%% file: ms.tex
\documentclass[structabstract]{aa}  
\usepackage{graphicx}
\usepackage{txfonts}
\usepackage{natbib}
\usepackage{longtable,lscape}
\def\teff{\mbox{$T_{\rm eff}$}}

\def\ebv{\mbox{$E(4405-5495)$}}
\def\rv{\mbox{$R_{5495}$}}
\def\logd{\mbox{$\log d$}}
\def\Al{\mbox{$A(\lambda)$}}
\def\al{\mbox{$a(\lambda)$}}
\def\AV{\mbox{$A_V$}}
\def\AVJ{\mbox{$A_{V_J}$}}
\def\ABJ{\mbox{$A_{B_J}$}}
\def\mum1{\mbox{$\mu$m$^{-1}$}}
\def\chir{\mbox{$\chi^2_{\rm red}$}}
\newcommand{\UJ}{\mbox{$U_J$}}
\newcommand{\BJ}{\mbox{$B_J$}}
\newcommand{\VJ}{\mbox{$V_J$}}
\newcommand{\BT}{\mbox{$B_T$}}
\newcommand{\VT}{\mbox{$V_T$}}
\newcommand{\uS}{\mbox{$u_S$}}
\newcommand{\vS}{\mbox{$v_S$}}
\newcommand{\bS}{\mbox{$b_S$}}
\newcommand{\yS}{\mbox{$y_S$}}
\newcommand{\JT}{\mbox{$J$}}
\newcommand{\HT}{\mbox{$H$}}
\newcommand{\KT}{\mbox{$K_{\rm s}$}}
\newcommand{\GG}{\mbox{$G$}}
\def\HII{\mbox{H\,{\sc ii}}}
\begin{document}

   \title{Optical-NIR dust extinction towards Galactic O stars}


   \author{J. Ma{\'\i}z Apell{\'a}niz\inst{1}
                  \and
                  R. H. Barb\'a\inst{2}
          }

   \institute{Centro de Astrobiolog{\'\i}a, CSIC-INTA, campus ESAC, camino bajo del castillo s/n, E-28\,692 Villanueva de la Ca\~nada, Spain \\
              \email{jmaiz@cab.inta-csic.es} \\
         \and
              Departamento de F{\'\i}sica y Astronom{\'\i}a, Universidad de La Serena, Av. Cisternas 1200 Norte, La Serena, Chile \\
             }

   \date{Received 5 October 2017; accepted 25 December 2017}

 
  \abstract
  {O stars are excellent tracers of the intervening ISM because of their high luminosity, blue intrinsic SED, and relatively featureless spectra. We are currently 
   conducting the Galactic O-Star Spectroscopic Survey (GOSSS), which is generating a large sample of O stars with accurate spectral types within several kpc 
   of the Sun.}
  {We aim to obtain a global picture of the properties of dust extinction in the solar neighborhood based on optical-NIR photometry of O stars with accurate spectral types.}
  {We have processed a carefully selected photometric set with the CHORIZOS code to measure the amount [\ebv] and type [\rv] of extinction towards 562 O-type 
   stellar systems. We have tested three different families of extinction laws and analyzed our results with the help of additional archival data. }
  {The Ma\'{\i}z Apell\'aniz et al. (2014, A\&A 564, A63) family of extinction laws provides a better description of Galactic dust that either the Cardelli et 
   al. (1989, ApJ 345, 245) or Fitzpatrick (1999, PASP 111, 63) families, so it should be preferentially used when analyzing samples similar to the one in this 
   paper. In many cases O stars and late-type stars experience similar amounts of extinction 
   at similar distances but some O stars are located close to the molecular clouds left over from their births and have larger extinctions than the average for
   nearby late-type populations. In qualitative terms, O stars experience a more diverse extinction than late-type stars, as some are affected by the
   small-grain-size, low-\rv\ effect of molecular clouds and others by the large-grain-size, high-\rv\ effect of \HII\ regions. Late-type stars experience a 
   narrower range of grain sizes or \rv, as their extinction is predominantly caused by the average, diffuse ISM. We propose that the reason for the existence
   of large-grain-size, high-\rv\ regions in the ISM in the form of \HII\ regions and hot-gas bubbles is the selective destruction of small dust grains by
   EUV photons and possibly by thermal sputtering by atoms or ions.}
  {}

   \keywords{Dust, extinction --- 
             Galaxy: structure ---
             Methods: data analysis ---
             Methods: observational ---
             Stars: early-type}

   \maketitle
%

\section{Introduction}

$\,\!$ \indent Soon after the discovery of the existence of an intervening interstellar medium (ISM) that obscures and reddens starlight, \citet{BaadMink37} realized 
that sightlines could differ not only in the amount of obscuration but also in its dependence with wavelength. In current terms, we say that the extinction law is not 
constant so in order to determine how the ISM changes the light we receive from the stars we need to specify both the amount and the type of extinction. Extinction is a
combination of absorption and scattering by the intervening particles and the largest contributor is dust, which produces mostly continuum extinction but also some
broad features in the UV and the IR. Atoms and molecules also have an effect in the form of discrete absorption lines and some absorption features (diffuse 
interstellar bands or DIBs) of unknown origin are also detected, but those will be mostly ignored in this paper, which deals with dust extinction.

Some astronomers are interested in dust extinction because of the information it contains regarding the spatial distribution and properties of the ISM but to most of them
it is simply a nuisance, an effect that complicates the calculation of luminosities, and all they want to know is how to eliminate it from their data. Throughout much
of the twentieth century astronomers have tried to generate extinction laws but kept hitting a wall: as different sightlines had different types of extinction, an
average extinction law could only be an approximation and one that was bound to fail miserably under some circumstances. It was not until the seminal work of 
\citet{Cardetal89} (CCM hereafter), that a family (not an average) of extinction laws was produced, with a parameter (\rv, see Appendix A for its relationship with
$R_V$) that characterized the type of extinction and that is associated with the average dust grain size (small grains yield a low value of \rv, large grains a high one).
CCM became the standard reference in the field for the next quarter of a century, as for the first time it allowed to simultaneously determine the amount and type of 
dust extinction and eliminate its effect from the observed photometry. CCM had a few problems, though, some of which were treated by the alternative family of extinction laws
of \citet{Fitz99} (F99 hereafter), and later on by the \citet{Maizetal14a} family, from now on MA14. MA14 showed that its family of extinction laws provided a better fit 
to the optical-NIR extinction in 30 Doradus than either CCM or F99, to the point that it was possible for the first time 
to obtain reasonable estimates of the effective temperatures of
O stars from photometry alone, something that was not possible until then due to the problems that exist in the other families. 
The MA14 family of extinction laws was derived using O stars on 30 Doradus so an obvious follow-up question is: does it also do a better job in characterizing optical-NIR
extinction for Galactic O stars? Answering that question is the first objective of this paper. 

In the last decade we have been conducting the Galactic O-Star Spectroscopic Survey or GOSSS \citep{Maizetal11}, which aims to obtain high S/N, $R\sim 2500$ blue-violet
spectroscopy of all optically accessible Galactic O stars and derive their spectral types. The sample 
currently available is the largest ever uniform collection of O spectral types
ever published. In this paper we use this sample to answer the question in the previous paragraph
and a second question: what is the relationship between dust grain size
and ISM environment? Previous attempts to answer this question have been hampered by, among other things, limited samples and the methodology employed. 
Our uses of the GOSSS sample and of the CHORIZOS \citep{Maiz04c} techniques applied in MA14 and in other recent works address both objectives.

This paper is divided as follows. We first present the sample in detail, describing how we have culled the photometry for the sample and eliminated problematic objects, and the
CHORIZOS methodology. We then show our results in three parts: a comparison of families of extinction laws, a discussion of quantitative reddening differences between O- and 
late-type stars, and an analysis of how the properties of dust are a function of the ISM environment. In the final standard section of the paper we present our conclusions and 
outline future lines of work. There are also 
five
appendices that discuss extinction terms and techniques in order to better understand the uncertainties and biases involved
in this type of analysis. We recommend readers who are unfamiliar with the mathematical aspects of extinction effects to start with those appendices. 

\section{Data and methods}

$\,\!$ \indent The sample in this paper is built starting from the 590 O stars in GOSSS-I \citep{Sotaetal11a}, GOSSS-II \citep{Sotaetal14}, and GOSSS-III
\citep{Maizetal16}\footnote{The GOSSS spectral types were derived from spectra gathered with five facilities: 
the 1.5~m Telescope at the Observatorio de Sierra Nevada (OSN), 
the 2.5~m du Pont Telescope at Las Campanas Observatory (LCO), 
the 3.5~m Telescope at Calar Alto Observatory (CAHA), 
and the 4.2~m William Herschel Telescope (WHT) and 10.4~m Gran Telescopio Canarias (GTC) at Observatorio del Roque de los Muchachos (ORM).}. 
For the initial sample we collected our {\bf basic photometric set}:

\begin{itemize}
 \item Ground-based Johnson \UJ \BJ \VJ\ photometry using Simbad \citep{Wengetal00} and the GCPD \citep{Mermetal97}.
 \item Near-infrared \JT \HT \KT\ photometry from 2MASS \citep{Skruetal06}.
 \item Optical \GG\ photometry from the Gaia Data Release 1 or DR1 \citep{vanLetal17}.
\end{itemize}
 
The majority of the stars in our sample have photometry in all seven of the basic photometric bands, though our quality control (see below) led us to 
eliminate some of them. We also compiled the {\bf extended photometric set}:

\begin{itemize}
 \item Tycho-2 \BT \VT\ photometry \citep{Hogetal00a}. 
 \item Str\"omgren \uS \vS \bS \yS\ photometry from \citet{Paun15}.
\end{itemize}

The reason to differentiate between basic and extended photometric sets is that in the final sample all objects include at least five of the seven basic 
photometric points but not necessarily the extended photometric ones, whose coverage is more limited (complete or near-complete for bright stars but 
incomplete for dim ones). This led us to divide the final sample into four subsamples: J2G, J2GT, J2GS, and J2GTS, where the naming convention refers to the 
photometry included, that is, J(ohnson), 2(MASS), G(aia), S(tr\"omgren), and T(ycho-2), respectively. 

After compiling the photometry for the initial sample we carefully analyzed each object to produce a final clean sample in the following way:

\begin{itemize}
 \item For the GOSSS objects with nearby companions (visual binaries and multiples) we analyzed which ones had separate Johnson, 2MASS, and Str\"omgren magnitudes
       and for which ones the available basic photometry was a combination of two or more components (because we had been able to obtain spatially resolved
       spectral types). In the latter case we merged the information on the different GOSSS components (see below for Gaia and Tycho-2). We also eliminated a few 
       cases where no reliable photometry could be found. This process reduced the sample to 571 stars. 
 \item The 2MASS photometry for bright stars is of very poor quality (uncertainties larger than 0.2 mag). Whenever possible, we searched for alternative 
       sources such as \citet{Duca02} to be used as substitutes. In the majority of cases where we used the 2MASS photometry we applied the 2MASS
       uncertainties directly, with the exception being the stars with poor quality flags, where we increased the uncertainty values. 
 \item For objects with NIR excesses we followed two different strategies. Oe stars, whose circumstellar environment can contribute significantly to the NIR 
       photometry, were eliminated, leaving a final sample of 562 stars. For the rest,
       we applied a correction to the \JT \HT \KT\ photometry based on the infrared colors
       that also increased the photometric uncertainties in those cases. The correction was based on the facts that [a] all O stars have very similar 
       photospheric intrinsic $\JT-\HT$ and $\HT-\KT$ colors and that [b] the extinction vector in the color-color plane has a nearly constant direction for 
       the range of \ebv\ values considered here (neither of those assumptions is true for most optical or optical-NIR colors, see  Appendix B). There are 
       numerous literature examples of this effect, one is shown in Fig.~4 of \citet{Ariaetal06}.
 \item For the values of the Johnson and Str\"omgren uncertainties, see \citet{Maiz06a}. In those cases where the source was suspected to be of poor quality,
       we increased the value of the uncertainties. For the zero points, see \citet{Maiz07a}.
 \item Tycho-2 photometry supposedly has no saturation limit while \GG\ saturates around magnitude 6.0 for DR1 \citep{vanLetal17,Maiz17}. However, when comparing
       photometry from different sources we discovered a slight saturation effect for the brightest Tycho-2 stars, so we eliminated those magnitudes from our
       analysis. 
       Regarding saturated \GG\ DR1 magnitudes, we used the \citet{Maiz17} correction.
 \item Gaia DR1 does not include variability information and \GG\ magnitudes appear to be the average ones. This effect turned out to be especially important for
       some eclipsing binaries, for which we also eliminated that magnitude.
       We used the calibration of \citet{Maiz17}, who corrected the nominal passband of \citet{Jordetal10}.
 \item Another relevant effect when comparing ground-based (Johnson, 2MASS, and Str\"omgren) and space (Gaia and Tycho-2) photometry is spatial resolution: the
       published space photometry may include different components of a multiple system that are seen as a single source from the ground. We revised all cases in turn to decide which components were detected. This led in some cases to combining the space photometry from two components (as we did with the
       ground-based photometry) and in others to discarding the Gaia or Tycho-2 photometric points.
 \item Finally, all the photometry was revised to determine the mutual compatibility in an iterative process that took weeks of work and led to a final culling
       of the used magnitudes. This is necessary because published photometry contains a significant number of errors
       \footnote{
       Variability is also an issue but a minor one. Most
       sources of variability in O stars such as the rotation of oblique magnetic fields \citep{Wadeetal12b}, wind, or
       turbulence variability significantly affect emission lines but produce only small changes in the continuum. Eruptive Oe stars 
       such as HD~120\,678 \citep{Gameetal12} have been excluded from the sample, so they should not concern us. The most relevant source of large broad-band flux 
       variability in the sample is eclipsing binaries. Most of them, however, are accounted for in the literature and those that are not can be revealed through 
       comparison of different sources. Nevertheless, it is possible that a few hidden binaries may be lurking in our sample, especially in the J2G subsample, 
       where there are less bands to compare among.}.
       The reader is invited to look up in online tools such as Simbad 
       and VizieR any of the stars in the sample with a large number of entries there. A quick perusal easily reveals cases in which, for example, Johnson photometry from
       different sources is incompatible. When comparing different photometric systems such incompatibility may not be obvious at first, as any two magnitudes 
       generate a color (even $\VJ-\yS$, something commonly ignored), but it becomes apparent when no valid single-star SED can be built with the combination
       independently of the parameters or extinction-law families used. We have recently started a different project called GALANTE (see below) with which we will 
       obtain uniform-quality photometry for a significant fraction of the sources here but its results are still far in the future.
\end{itemize}

\begin{figure}
\centerline{\includegraphics[width=\linewidth]{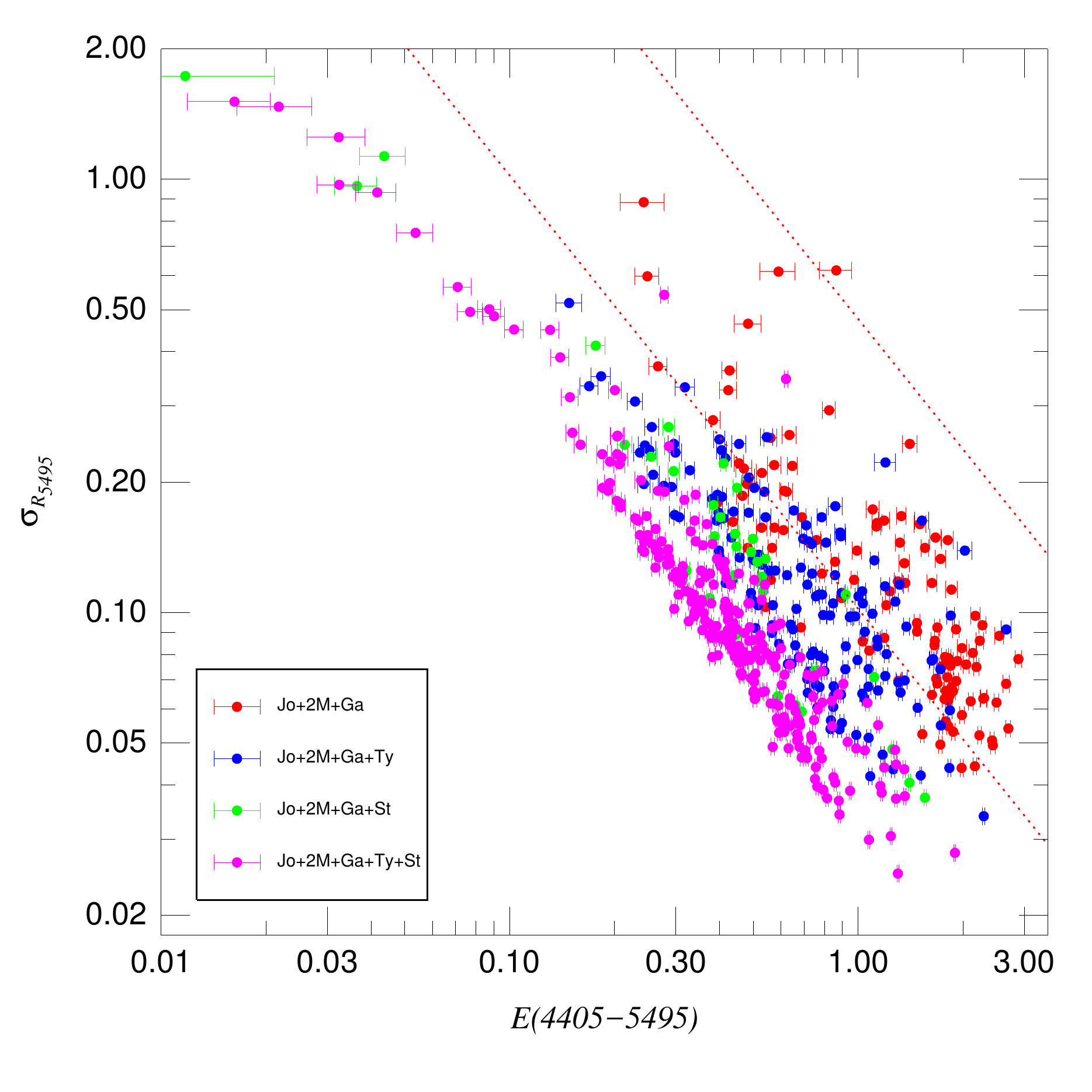}}
\caption{$\sigma_{R_{5495}}$ as a function of \ebv\ for the MA14 extinction family CHORIZOS runs. Colors are used to distinguish the subsamples: red for 
         J2G, blue for J2GT, green for J2GS, and magenta for J2GTS. The two dotted lines show the prediction of Eq.~\ref{sigmarv} for 
         [a] $\sigma_{A_V}$ = 0.03 and \rv\ = 3.0 (lower line) and [b] $\sigma_{A_V}$ = 0.10 and \rv\ = 5.0 (upper line).}
\label{ebv_srv}
\end{figure}

The collected Johnson photometry is available from the web site of the Galactic O-Star Catalog (GOSC, \citealt{Maizetal04b,Sotaetal08})
\footnote{We have recently moved the main URL to {\tt https://gosc.cab.inta-csic.es} from the old URL at {\tt http://gosc.iaa.es}, which will be kept as a
mirror site for the time being.}.
The final subsamples have 109 (J2G), 152 (J2GT), 38 (J2GS), and 263 (J2GTS) stars. Of the final 562 stars, 28, 6, 13, 13, and 
111
have \UJ, \JT, \HT, \KT, and \GG\ missing, respectively.

The photometry was processed using the latest version of the CHORIZOS code \citep{Maiz04c} to determine the amount and type of extinction to each of the star. We 
have used this method successfully in the past (e.g., \citealt{Maizetal15a}), here we describe the specific details of the CHORIZOS runs used in this paper:

\begin{itemize}
 \item We used the Milky Way grid of \citet{Maiz13a}, in which the two grid parameters are effective temperature (\teff) and photometric luminosity class 
       (LC). The latter quantity is defined in an analogous way to the spectroscopic equivalent, but instead of being discrete it is a 
       continuous variable that ranges from 0.0 (highest luminosity for that \teff) to 5.5 (lowest luminosity for that \teff). We note that the range is 
       selected in order to make objects with spectroscopic luminosity class V (dwarfs) have LC$\approx$5 and objects with spectral luminosity class I
       (supergiants) have LC$\approx$1. In this paper we are only interested in O stars, whose spectral energy distributions (SEDs) are TLUSTY \citep{LanzHube03}.
 \item Each star was processed three times, once each with the CCM, F99, and MA14 families of
       extinction laws. The type of extinction is parameterized by \rv\ and the amount of extinction\footnote{Throughout this paper we
       refer to \ebv\ as the amount of extinction, though it should be more properly called the amount of reddening, because that is the choice the CCM, F99, and
       MA14 families use.} is parameterized by \ebv. See Appendix A for a discussion on the use of those parameters.
 \item The \teff-spectral type conversion used is an adapted version of \citet{Martetal05a} that includes the 
       spectral subtypes and luminosity classes used in GOSSS I+II+III.
 \item \teff\ and LC were fixed while \rv, \ebv, and \logd\ were left as free parameters. The values of the \teff\ were established from the used 
       \teff-spectral type conversion (see previous point). For LC we used the spectroscopic luminosity class available from the GOSSS spectral classification. 
       We note that, in any case, optical-NIR colors of O stars are very weakly dependent on luminosity other than wind effects.
 \item The four subsamples were run separately. The number of photometric points for the stars in each sample is 7 (J2G), 9 (J2GT), 11 (J2GS), and 13
       (J2GTS). As we are fitting three parameters with CHORIZOS, the degrees of freedom (d.o.f.), are four, six, eight, and ten, respectively\footnote{In those
       cases where some photometric point is missing, d.o.f. is reduced accordingly.}. The reduced $\chi^2$ of the best model, \chir = $\chi^2$/d.o.f. is used to 
       evaluate the quality of the fit for each star.
\end{itemize}

\input{maintable}

\begin{figure*}
\centerline{\includegraphics[width=0.49\linewidth]{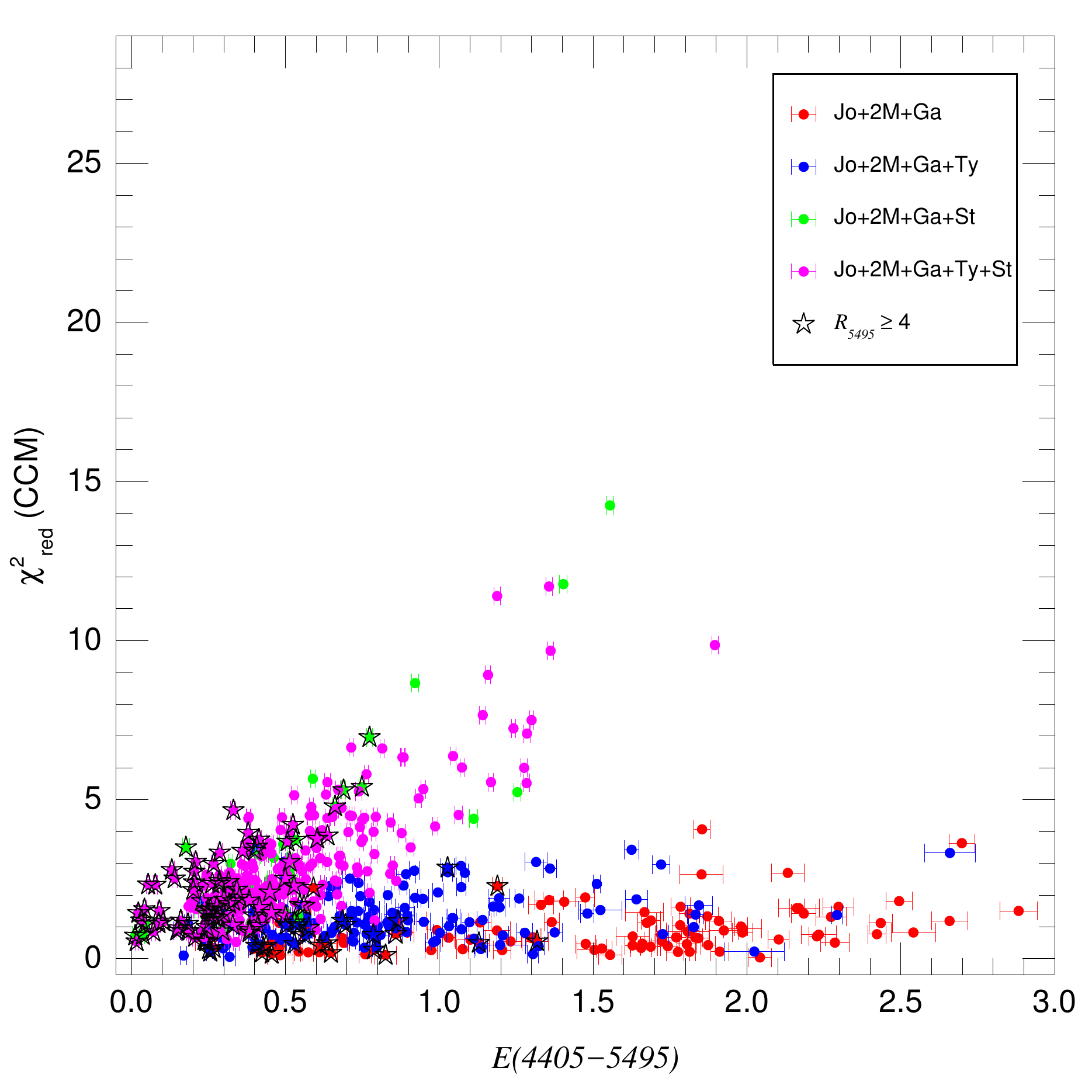} \
            \includegraphics[width=0.49\linewidth]{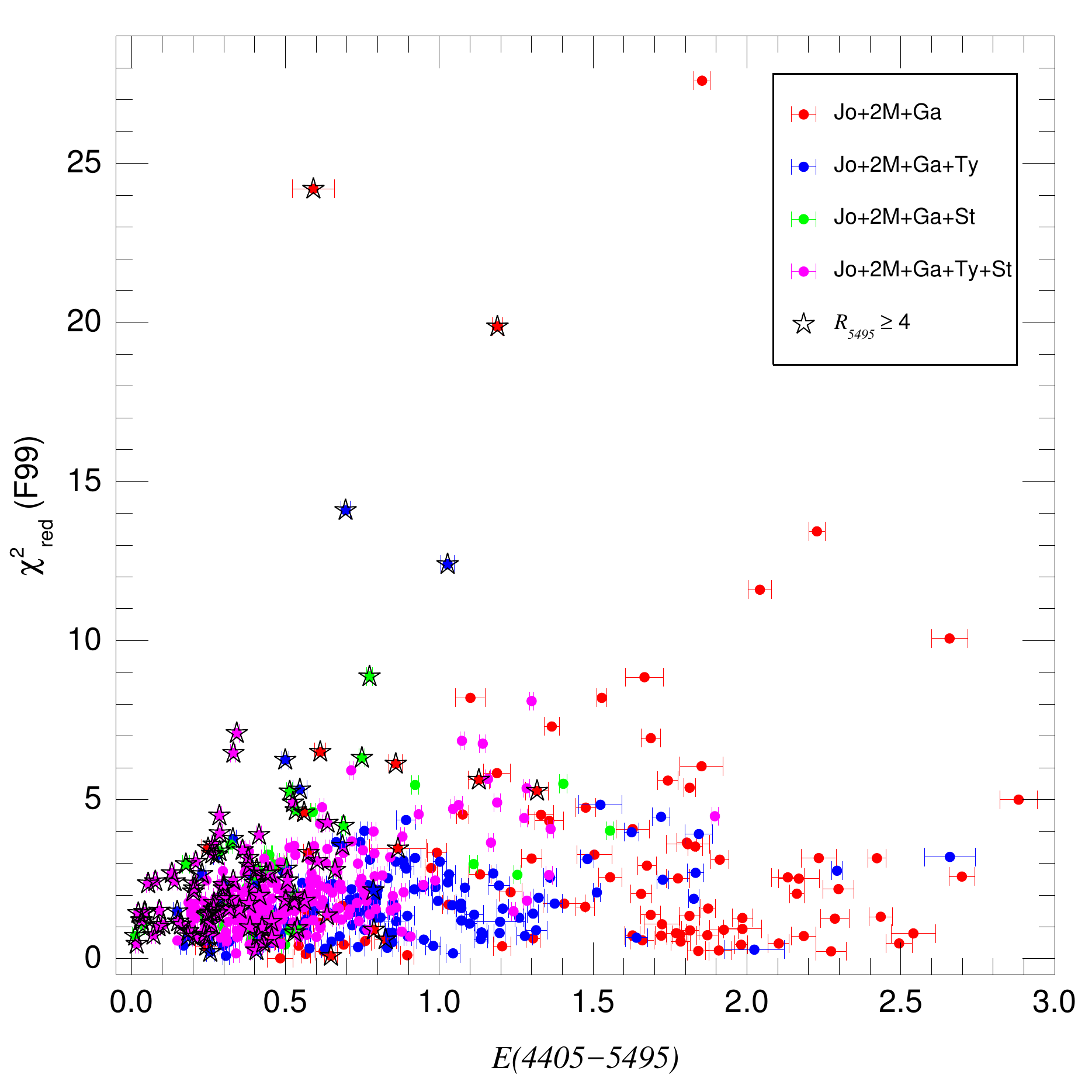}}
\centerline{\includegraphics[width=0.49\linewidth]{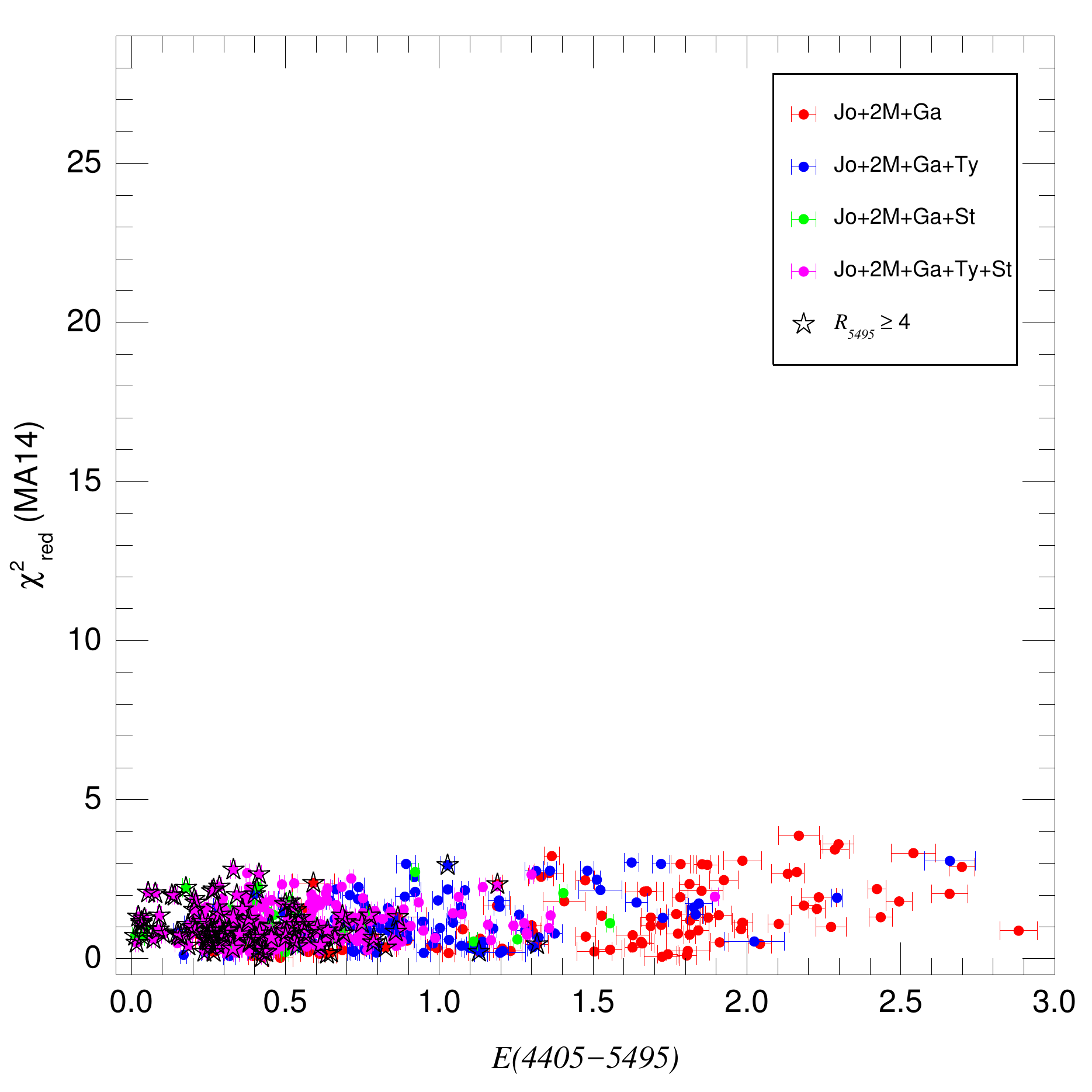}}
\caption{\chir\ as a function of \ebv\ for the CHORIZOS runs in this paper using the CCM (upper left), F99 (upper right), and MA14 (bottom) families of 
         extinction laws. The color coding is the same as in Fig.~\ref{ebv_srv}. In addition, objects with $\rv \ge 4$ are indicated with a black star symbol. 
         In all cases, the \ebv\ and \rv\ values are the MA14 results. Both axes scales are the same in the three plots to allow for a better comparison.}
\label{exlawscomp1}
\end{figure*}

\begin{figure*}
\centerline{\includegraphics[width=0.49\linewidth]{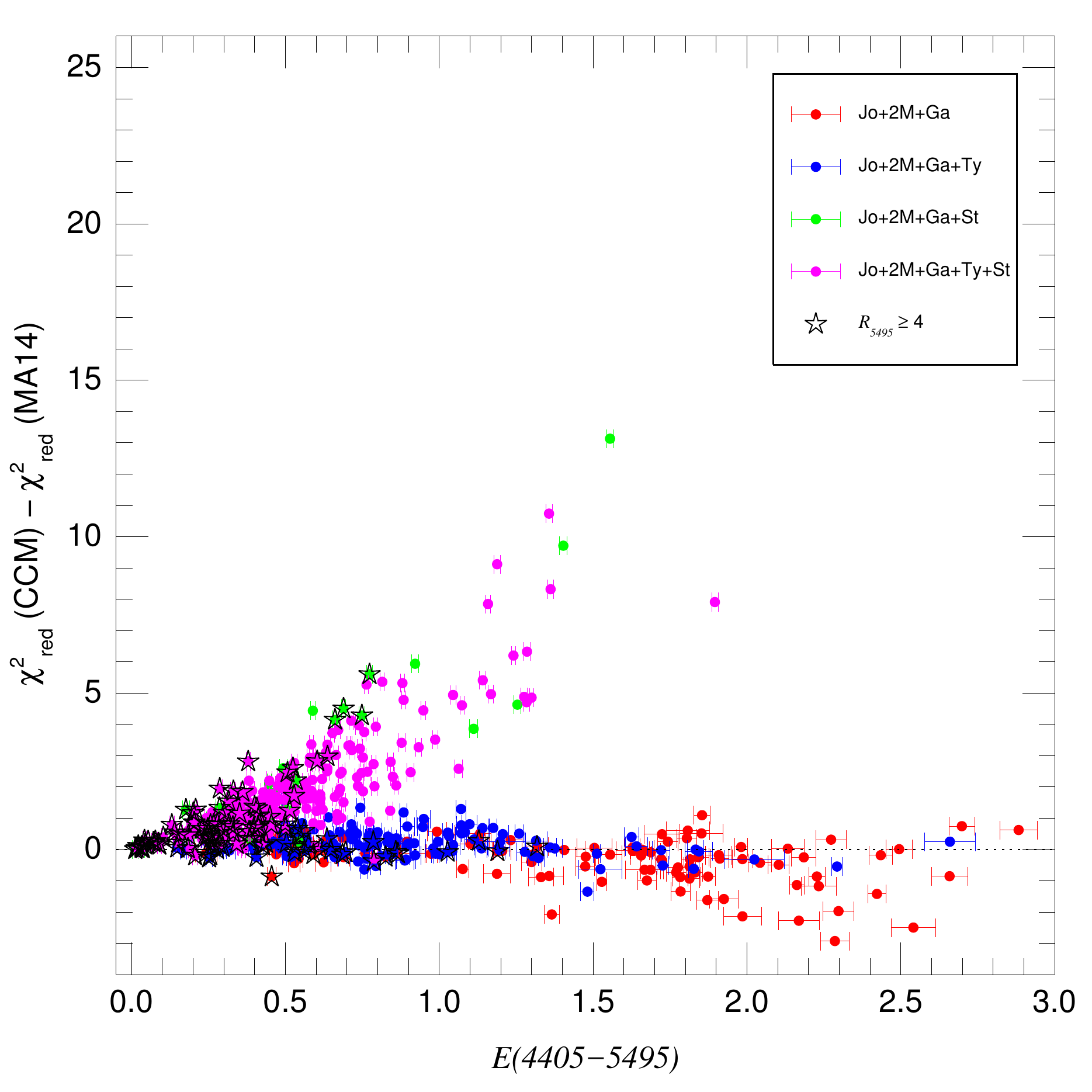} \
            \includegraphics[width=0.49\linewidth]{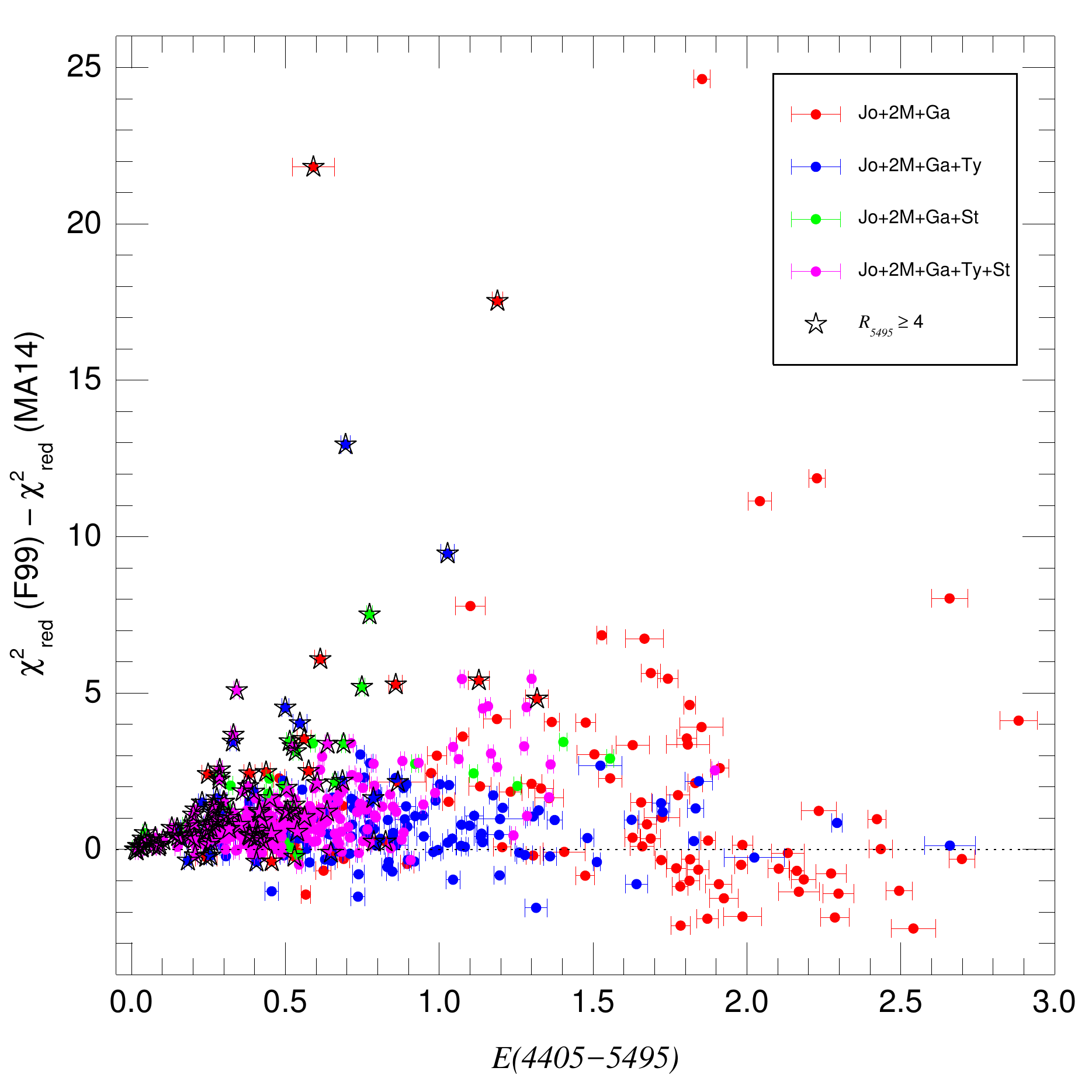}}
\caption{\chir\ difference between CCM (left) / F99 (right) and MA14 as a function of \ebv\ for the CHORIZOS runs in this paper. The color coding and symbols
         are the same as in Fig.~\ref{exlawscomp1}. In all cases, the \ebv\ and \rv\ values are the MA14 results. Both axes scales are the same in the two
         plots to allow for a better comparison.}
\label{exlawscomp2}
\end{figure*}

The results from the CHORIZOS runs are given in Table~\ref{maintable}, 
also available in electronic form at the CDS via anonymous ftp to {\tt cdsarc.u-strasbg.fr} (130.79.128.5) or via {\tt http://cdsweb.u-strasbg.fr/cgi-bin/qcat?J/A+A/}.
The spectroscopic logarithmic distances (\logd) are given with only one digit after 
the decimal point as that is their expected level of precision\footnote{Gaia is expected to provide more accurate trigonometric distances to most of the sample
in this paper in a short time scale but we note that the extinction parameters derived in this paper such as \ebv\ and \rv\ are only very weakly dependent on 
distance.}. In the next section we discuss the scientific results. Prior to that, we
comment on some issues regarding the methodology. The reader is referred to the Appendices for explanations of why we have used monochromatic quantities 
instead of filter-integrated ones to characterize extinction and on some of the numerical consequences of using multifilter photometry to determine extinction
properties.

The quantities listed in Table~\ref{maintable} are the direct output of CHORIZOS and the uncertainties should be taken as the random ones. They do
not include systematic uncertainties, of which there are three possible origins:

\begin{itemize}
 \item {\it Photometric.} Here we have possible zero-point errors and discrepancies between different photometric sources. As previously mentioned, 
       we have minimized those problems by analyzing the zero points in different systems \citep{Maiz05b,Maiz06a,Maiz07a} and by assigning quality flags
       (and associated uncertainties) to different sources.
 \item {\it SED.} In principle, the optical-NIR SEDs of O stars are well known and depend little on metallicity. However, one should be careful not to
       include or to correct for those cases where the SED is anomalous. We have done this by excluding Oe stars from our sample and by correcting for
       NIR excesses when present.
 \item {\it Extinction laws.} One can obtain results assuming any extinction law. However, that does not mean that the 
       used extinction law is the correct one. If it is not, a systematic error is likely to be introduced in the process. In the next section we discuss
       how we have checked this.
\end{itemize}

Regarding systematic uncertainties, we specifically point out the differences in the uncertainties in $V_{J,0}$ as calculated by two possible methods 
described in Appendix C. The uncertainties in Table~\ref{maintable} are calculated using the whole CHORIZOS likelihood grid (second method in Appendix C). 
As a consequence, the uncertainties in $V_{J,0}$ are smaller than those in \AVJ. For \logd\ we only give two significant figures and no random uncertainties, 
as there are two large sources of systematic uncertainties: the intrinsic width of the luminosity class-luminosity relationship and the non-inclusion of the 
existence of multiple systems (astrometric or spectroscopic) in the photometric data. The reader is referred to \citet{Maizetal15c,Maizetal15a} for an example 
of how those issues complicate the use of spectroscopic parallaxes for O stars. We will revisit this issue in future papers once accurate Gaia parallaxes are 
available for a large fraction of the sample.

Figure~\ref{ebv_srv} shows the dependence of the \rv\ uncertainties from the CHORIZOS runs on \ebv. For the J2 sample the values agree reasonably well
with the predictions of Eq.~\ref{sigmarv}. The results for the other subsamples indicate that it is possible to reduce the random uncertainty of \rv\
by including additional filters in the analysis, that is, the more (good) data, the better,
as expected.

\section{Results}

\subsection{Comparing extinction-law families}

\begin{figure}
\centerline{\includegraphics[width=\linewidth]{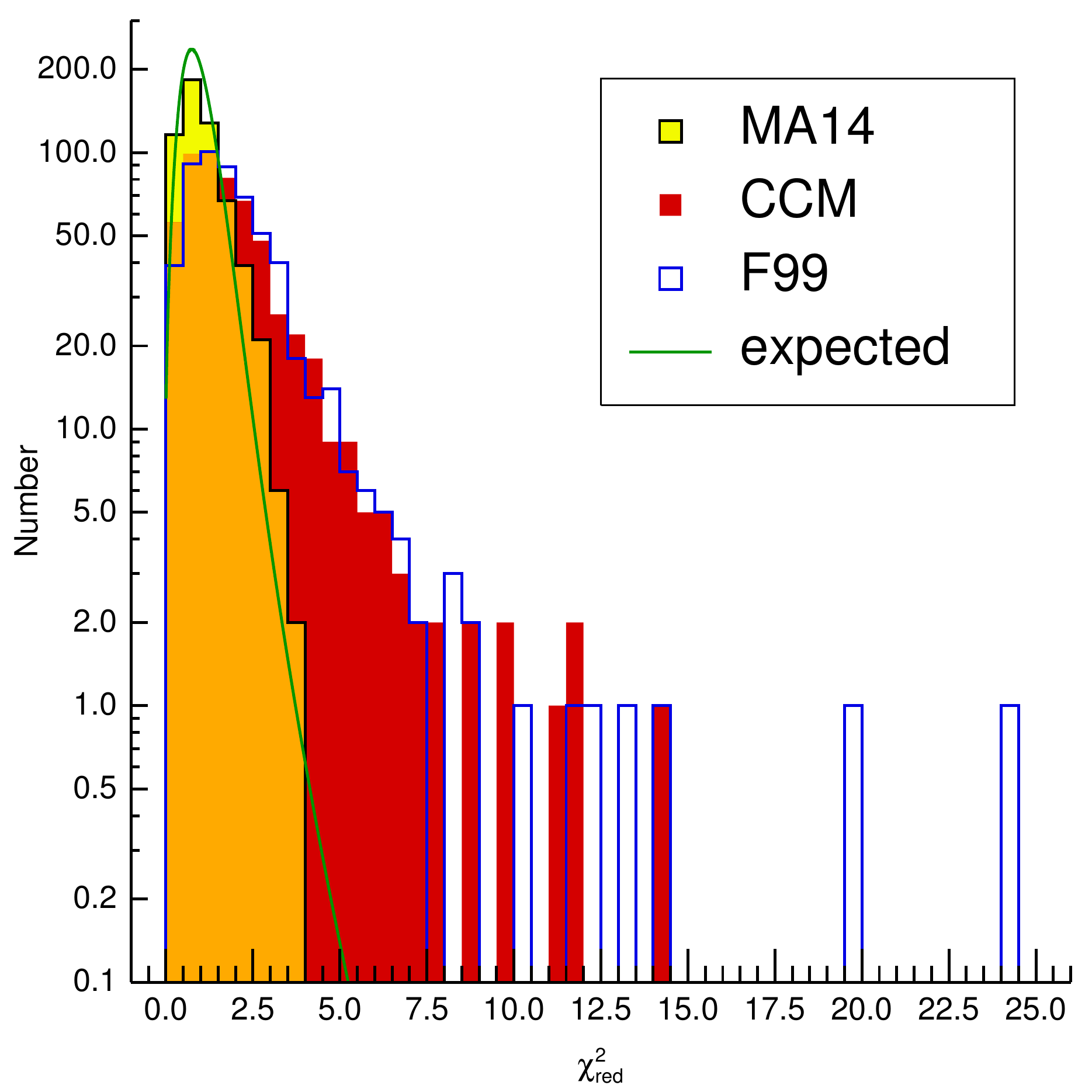}}
\caption{\chir\ histograms for the CHORIZOS runs in this paper using the MA14, CCM, and F99 families of extinction laws. We also plot the expected combined
         distribution. We note that the orange region represents the superposition of the yellow (MA14) and red (CCM) histograms.}
\label{chi2stats}
\end{figure}

$\,\!$\indent We begin our analysis by comparing the results obtained with the CCM, F99, and MA14 families of extinction laws
\footnote{We refer the reader to Appendix~E for an explanation of why we did not include \citet{FitzMass09} 
in our list of comparison families of extinction laws.}.
In Fig.~\ref{exlawscomp1}
we plot \chir\ as a function of \ebv\ for each family. Figure~\ref{exlawscomp2} shows two similar plots but with the \chir\ differences between MA14 and
CCM (left) / F99 (right)
on the vertical axis. Figure~\ref{chi2stats} shows the \chir\ histograms for the three families compared with the expected combined 
distribution, built from the sum of the distributions for each star, each one with its d.o.f.

The most important result of the comparison is that the MA14 family provides much better results than either CCM or F99. For MA14 all stars have 
$\chir < 4.0$ while for the other two families there is a long tail that extends to values above ten. The MA14 results show no strong trends as a
function of \ebv\ or differences between subsamples. Its \chir\ histogram has a cutoff at the expected location but is more heavily populated in the
1.5-3.0 region than the expected distribution. This could be caused either by systematic errors in the input photometry (for example, due to variability)
or by the need to fine-tune 
the extinction laws. However, that issue appears to be a minor effect for low-intermediate extinction values, so we can conclude
that {\bf the MA14 family of extinction laws provides the best description of Galactic optical extinction for $\ebv < 3.0$ available to date.} 
This statement is true even though the MA14 laws were derived using 30 Doradus, not Galactic, data; such validity was already noted in 
the MA14 paper itself for a very limited sample of Galactic stars. We also note that we had already reached this conclusion when we presented the
preliminary results of this paper at two scientific conferences \citep{Maiz15a,Maizetal17b}. 

\begin{figure}
\centerline{\includegraphics[width=\linewidth]{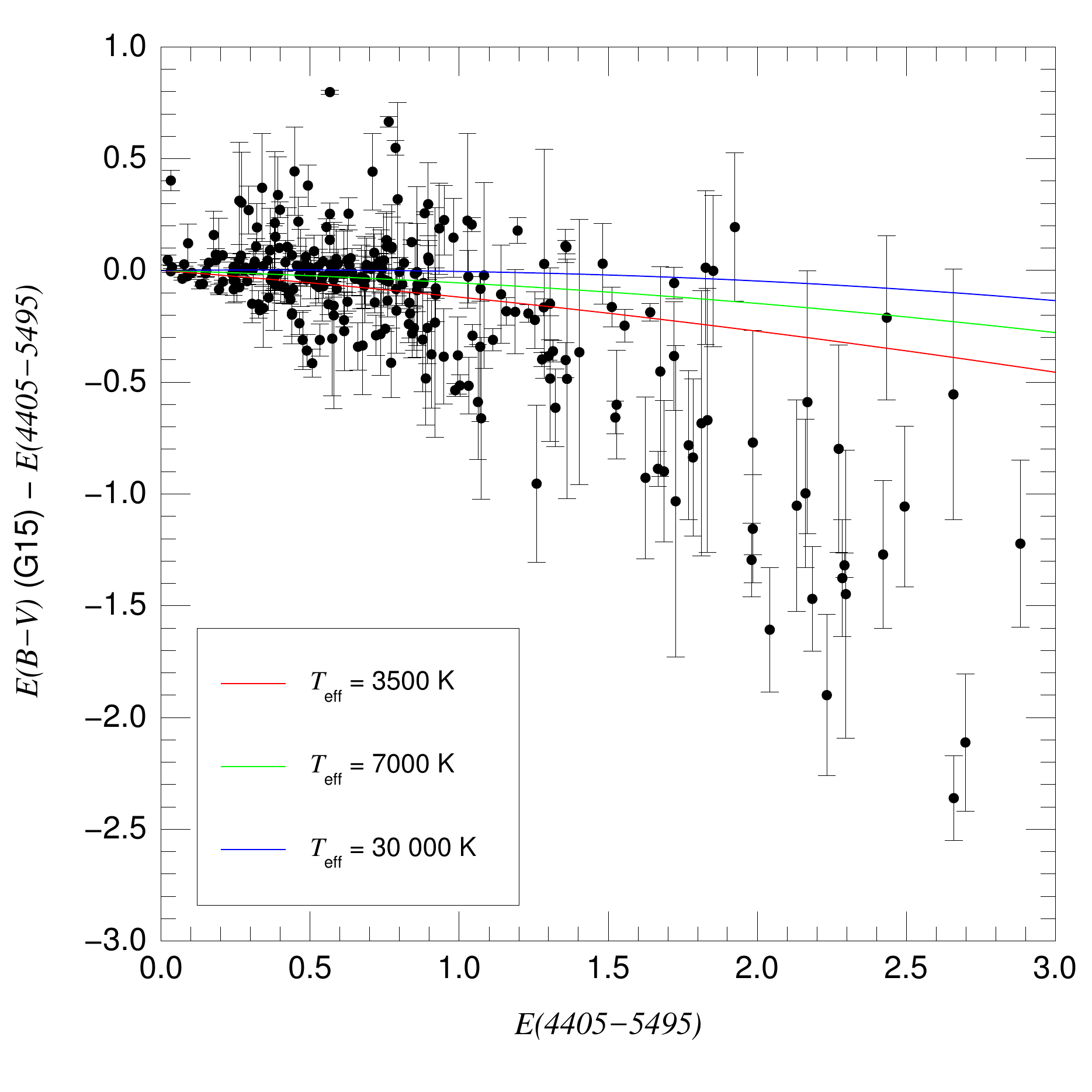}}
 \caption{Comparison between the reddenings measured by G15 and those in this paper. The error bars include the uncertainty in \logd\ (G15 gives $E(B-V)$ as
          a function of distance), assumed to be 0.1 (corresponding to an uncertainty of 0.5 in the distance modulus). The three colored lines show the expected 
          difference between $E(B-V)$ and \ebv\ for three different temperatures assuming an MA14 extinction law with \rv\ = 3.0.}
\label{argonaut}
\end{figure}

Why does the MA14 family work better than either CCM or F99? In the case of CCM there are two culprits. the use of a seventh-degree 
polynomial in $1/\lambda$ for wavelength interpolation \citep{Maiz13b} and the behavior for the $U$ band (see MA14). The first effect is clearly seen in the left
panels of Figs.~\ref{exlawscomp1}~and~~\ref{exlawscomp2}: for a given value of \ebv\ the typical \chir\ is worse when Str\"omgren photometry (which has  
filters intercalated between $U$ and $B$ and between $B$ and $V$) is present than when is not. The F99 results are significantly worse than for MA14. This is 
especially true for large values of \rv, which are the worst offenders for $\ebv < 1.5$ (as explained below, our sample does not contain objects with both large
\ebv\ and \rv).

It is interesting to compare our results with those of \citet{Schletal16}. Those authors find that the MA14 family provides a good description of the mean optical 
extinction curve and of its variation with \rv. However, they find that the quality of the MA14 fit becomes poorer in the NIR. The explanation for the latter effect 
is that their photometry 
has a better coverage for wavelengths longer than 6000~\AA\ (our optical photometry only has the redwards portion of Gaia \GG\ in that range) 
and an average reddening larger than the one in our sample: the MA14 derivation assumed the \citet{RiekLebo85} 
power-law form in the NIR with $\alpha = -1.61$ because of the impossibility of using the relatively low-extinction 30 Dor stars to measure $\alpha$. This is an 
illustration of a chronic problem in extinction studies: a strong extinction effect in the UV becomes a weak one in the optical and an undetectable one in the IR 
while a star with a strong effect in the IR is hard to observe in the optical and impossible in the UV. As a result, one is forced to build UV-optical-IR extinction 
laws by stitching together results for stars of different amount of extinction. In the specific case of MA14, that paper specifically mentions this is an expected 
issue and recommends not using its results for large NIR extinctions: that is also why we indicate above that the results of this paper apply to the optical for 
$\ebv < 3.0$.  For a more detailed discussion of the different wavelength regimes, see \citet{Maiz15b} and \citet{Maizetal17b} and the last subsection below. 

\subsection{Quantitative reddening differences between O-type and late-type stars}

\begin{figure*}
\centerline{\includegraphics[width=0.605\linewidth]{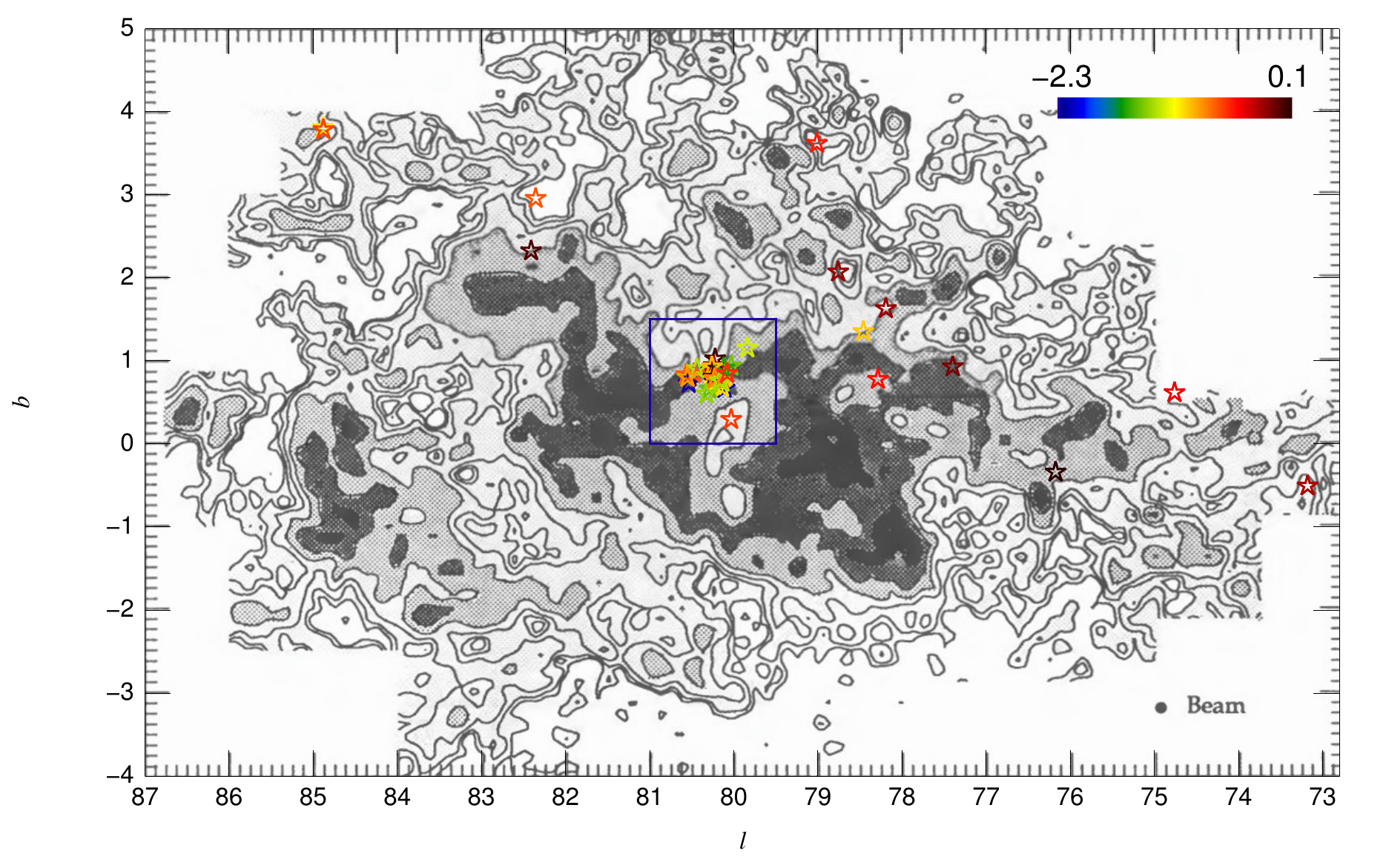} \
            \includegraphics[width=0.385\linewidth]{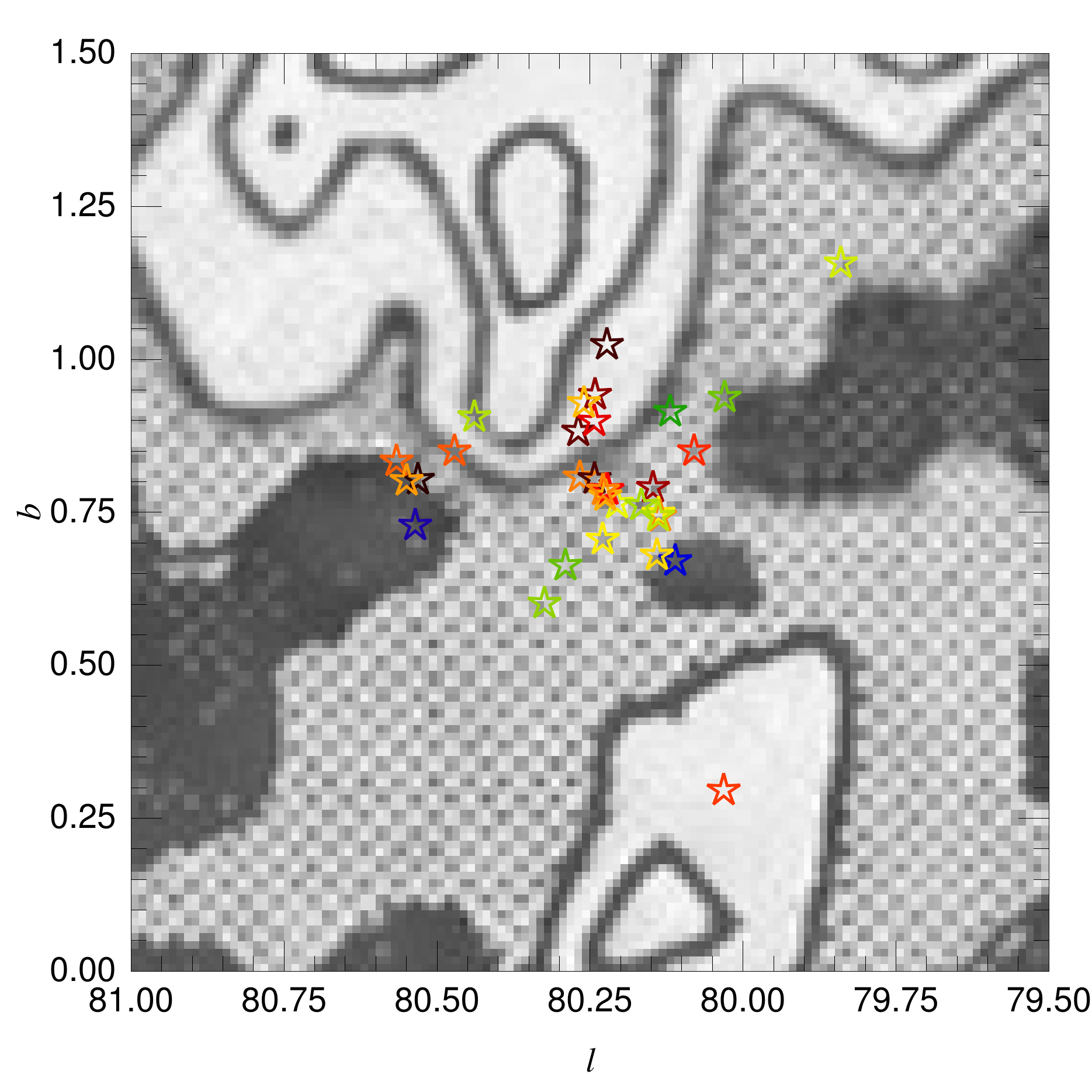}}
 \caption{Difference between the G15 $E(B-V)$ values and \ebv\ for the O stars with $\ebv > 1.0$ in the Cygnus-X region plotted on top of the CO $J = 1\leftarrow 0$ 
          map of \citet{LeunThad92}. The value is encoded in the color of each symbol (see scale on left panel and compare with the vertical axis of 
          Fig.~\ref{argonaut}). The left panel shows the whole region while the 
          right panel zooms in the blue rectangular region, which corresponds to Cyg OB2. The beam FWHM is 8\farcm7 ($\sim 4$ pc at the distance of Cyg OB2), so 
          the CO clouds are likely to have substructure not seen in the map. The angular resolution of the G15 values is similar to the beam size. The axes are 
          in Galactic coordinates and the units are degrees.}
\label{CygXCO}
\end{figure*}

\begin{figure*}
\centerline{\includegraphics[width=0.49\linewidth]{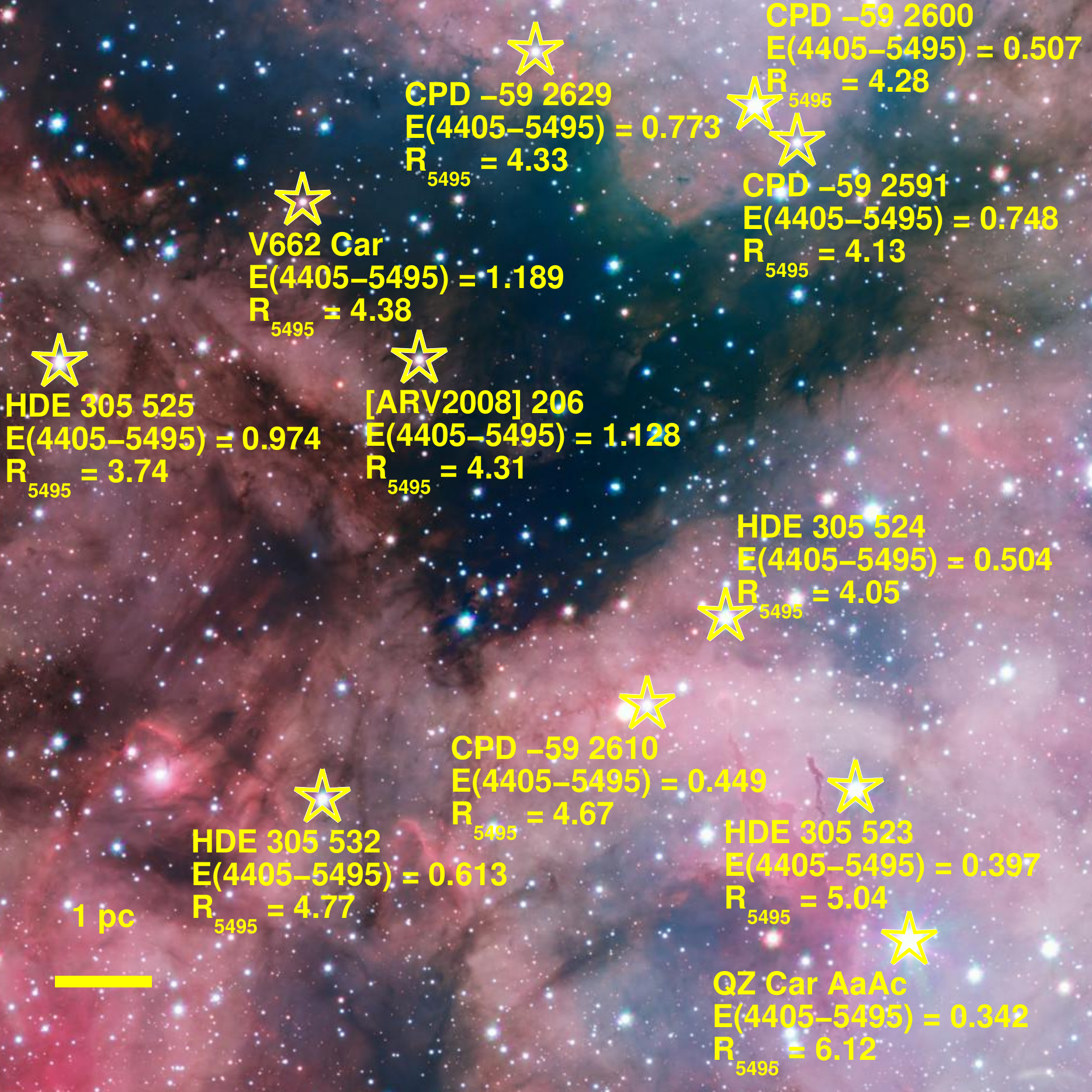} \
            \includegraphics[width=0.49\linewidth]{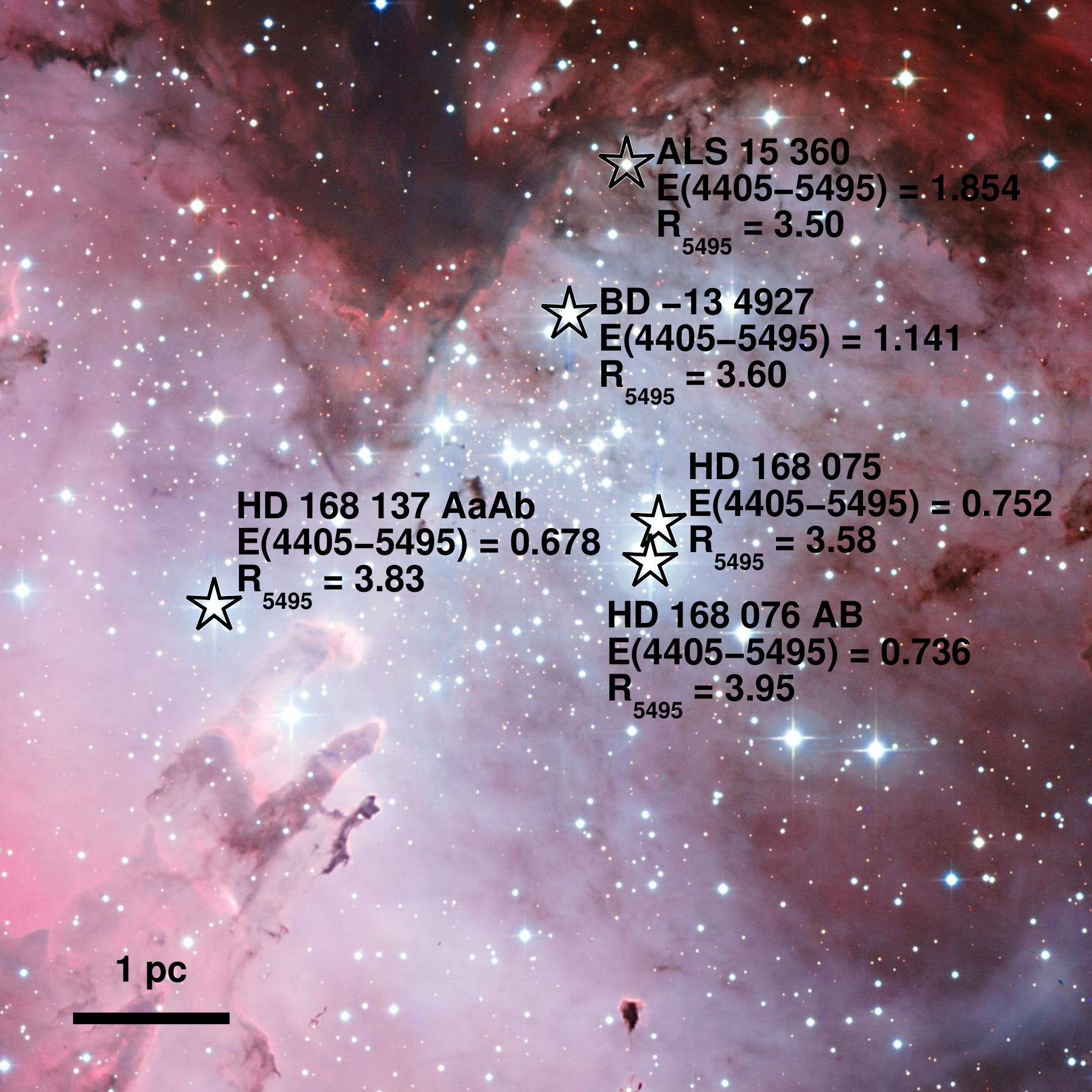}}
\smallskip
\centerline{\includegraphics[width=0.49\linewidth]{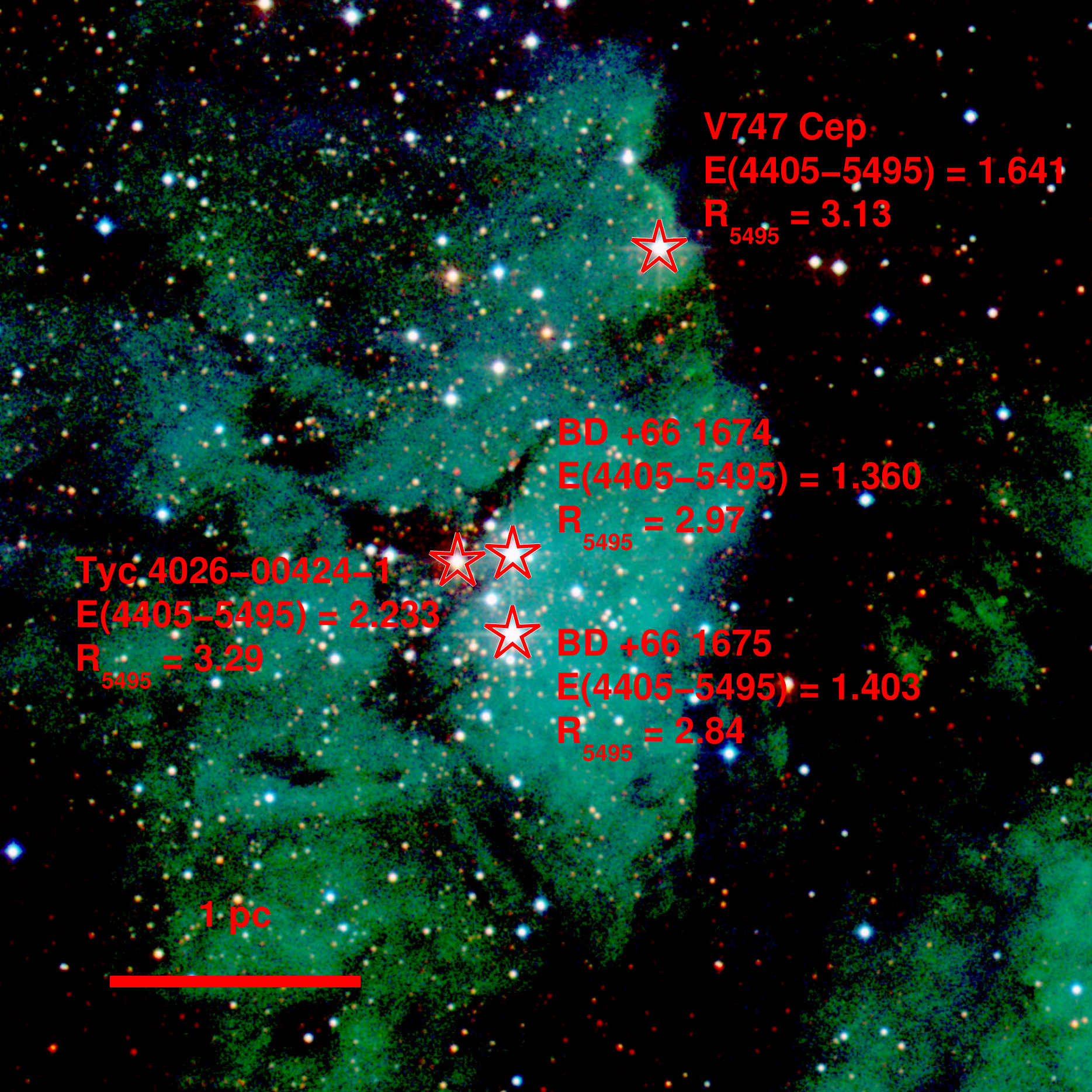} \
            \includegraphics[width=0.49\linewidth]{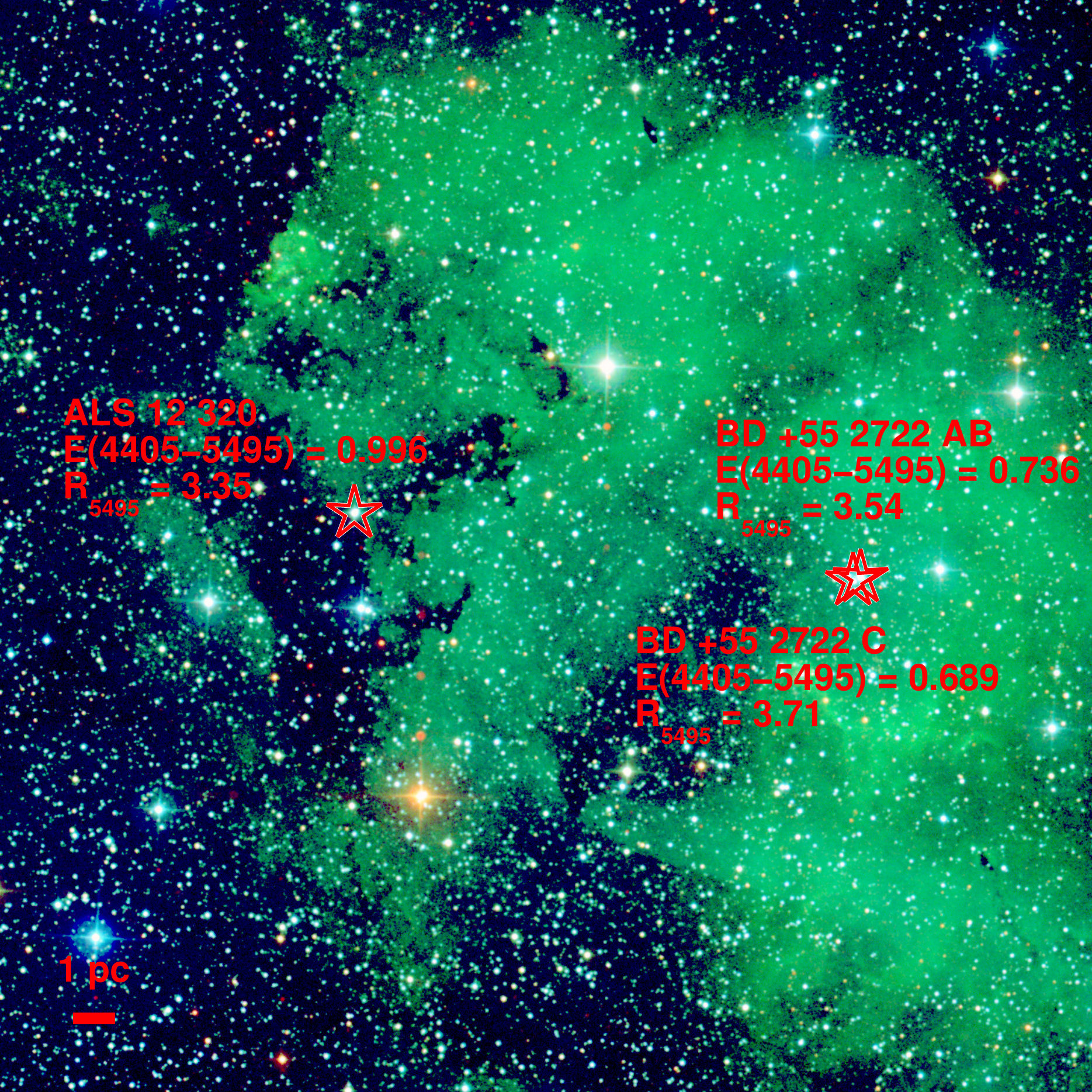}}
\caption{Extinction measurements for O stars in four \HII\ regions: (top left) the Carina Nebula, (top right) M16/NGC~6611, (bottom left) Berkeley~59/NGC~7822, and
         (bottom right) Sharpless~2-132. The top two images are from ESO press releases 1250 ($Bgr$H$\alpha$) and 0926 ($BVR$), respectively. The bottom two images
         are RGB combinations of $J_{\rm 2MASS}$+$R_{\rm DSS2}$+$B_{\rm DSS2}$. North is towards the top and east towards the left and the approximate physical 
         scale is indicated in all cases. Only a section of each \HII\ region is shown to better visualize the dust lanes.}
\label{HIIregions1}
\end{figure*}

$\,\!$\indent Older studies of the spatial distribution of extinction (e.g., \citealt{Fitz68,Necketal80}) were based on the analysis of early-type stars. Such studies 
had systematic uncertainties due to some of the issues discussed in this paper (photometric biases, extinction law errors, spectral type mismatches) and were limited
by small samples, as early-type stars are relatively scarce and fewer of them had accurate spectral types at the time of those studies compared to today. More
recent studies have taken advantage of larger photometric databases and concentrate on the more numerous late-type stars 
\citep{Marsetal06,Lalletal14,Saleetal14,Greeetal15}. 
The use of late-type stars allows for a much better spatial sampling that is allowed with early-type stars but that does not eliminate the need for early-type
studies because extinction may not affect different spectral types in the same manner (statistically speaking). Given the young age of 
O stars, they are expected to be located preferentially close to their natal molecular clouds. Also, their strong UV fluxes ionize the
surrounding ISM, in some cases producing H\,{\sc ii} regions. In either circumstance, either the quantity or the type of extinction they experience should be
different from the typical extinction experienced by late-type stars, which should be dominated by the diffuse Galactic ISM. In other words, {\bf the light we receive 
from O stars is expected to have crossed a more local, clumpy, and diverse ISM than the light from late-type stars}. This idea is what prompted us to study 
O-star extinction taking advantage of modern capabilities and data. In this subsection we analyze the quantitative aspects (amount of extinction) and in the following subsection
we analyze the qualitative ones (type of extinction).

To test the differences between the extinction experienced by O-type and cool stars we use the results from \citet{Greeetal15}, G15 from now on. G15 combined 
a large sample of Pan-STARRS 1 \citep{Kaisetal10} and 2MASS photometry from mostly cool stars to derive a 3-D reddening map of three quarters of the sky with an 
angular resolution of 3\farcm4-13\farcm7 and a maximum distance resolution of 25\%. Their database contains the 
$E(B-V)$ for 262 of the stars in our sample assuming the values of \logd\ from Table~\ref{maintable}. We compared them with our \ebv\ results in Fig.~\ref{argonaut}, 
where we also plot the expected relationship between $E(B-V)$ and \ebv\ for three different $T_{\rm eff}$ using an MA14 extinction law with \rv\ = 3.0 (see 
\citealt{Maiz13b} and Appendix A). For most low-extinction stars our results are consistent with those of G15: in those cases there appears to be nothing special 
regarding the ISM that surrounds the O stars, making their reddenings similar to those of late-type stars in their neighborhood.

On the other hand, for the majority of objects with $\ebv\ > 1.0$, G15 systematically and growingly underestimates O-star reddenings by approximately a factor of 2. 
This indicates that such O stars have an additional reddening component that is not present in the sightline towards the surrounding late-type stars. That additional 
component is likely to be the dense gas clumps left over from the natal cloud that are 
known to have sub-pc structures (see e.g., \citealt{Alveetal01,Scapetal02,Maizetal15c}). Therefore, O stars located at $\sim$1 pc projected distances
to the clouds will experience the additional extinction only in a statistical sense, as some sightlines will intercept the dense cores and some will not. Most 
late-type stars will be located farther away from the clouds and only a few of their sightlines will be affected by them. This interpretation is consistent with the 
scatter seen in Fig.~\ref{argonaut} for $\ebv\ > 1.0$.

The differences between the G15 $E(B-V)$ values and \ebv\ are clearly seen for the Cygnus-X region in Fig.~\ref{CygXCO}, where the most negative values take place at 
the location of the CO clouds with the highest column depths. However, the beam size is too large to visualize the correlation between the CO gas and the clumpy 
extinction at sub-pc scales. To do this, it is better to look at the relationship between the foreground dust lanes that obscure \HII\ regions in the optical and the 
amount of extinction affecting O stars. We have selected four Galactic \HII\ regions that have three or more O stars where the correlation between dust lanes and 
extinction is clear, and we discuss them with reference to Fig.~\ref{HIIregions1}:

\begin{itemize}
 \item The top left panel shows a section of the Carina Nebula, whose optical structure is dominated by a V-shaped dust lane several tens of pc in size (see e.g., 
       \citealt{Smitetal00a}). The section shown corresponds to the southern tip of the dust lane (in the top half of the image), where the two segments of the V meet,
       and where part of the dust lane is seen as a CO cloud \citep{Reboetal16}. Two stars, V662~Car (a.k.a. FO~15) and [ARV2008]~206 are located inside the dust 
       lane and those two objects are the ones with the highest reddening 
       [\ebv$\sim$1.1].
       On the other hand, the two objects located farther away from the dust lane, 
       HDE~305\,523 and QZ~Car~AaAc (a.k.a. HD~93\,206~AaAc) have the lowest reddening [\ebv$<$0.4]. The rest of the stars, located
       closer to the dust lane, have intermediate values of \ebv. Therefore, in this case it is clear that the dust lane is a large factor in the differential extinction
       in the Carina Nebula\footnote{We note that such an association may not be present in some cases, as it is possible for a star to be in the foreground with respect to
       the dust lane.}
 \item A section of M16, including the iconic ``pillars of creation'' \citep{Hestetal96} is shown in the top right panel. The famous 
       pillars themselves (in the bottom left quadrant) are at least partially immersed in the \HII\ region, as indicated by their surface brightness in emission lines, 
       but a bigger dark cloud towards the north that includes a thicker pillar (see \citealt{Hilletal12}) is likely to be in the foreground and is partially 
       responsible for giving the nebula its eagle shape. There are three objects from our sample near the center of the nebula (HD~168\,137~AaAb, HD~168\,076~AB, 
       and HD~168\,075) with \ebv$\sim$0.7. The two objects closer to the thick northern pillar have significantly higher reddenings, indicating that the dark cloud is 
       responsible for the additional extinction experienced by them (see \citealt{Belietal99}). 
 \item NGC~7822 is an \HII\ region ionized by the Berkeley~59 cluster, as shown in the bottom left panel. Three O stars, BD~+66~1674, BD~+66~1675, and 
       Tyc~4026-00424-1 are located near the center of the image and of the H$\alpha$ nebulosity. The first two have very similar extinctions but the third one is
       located at the same position in the sky as a dust lane and that increases its reddening by one magnitude. A fourth star, V747~Cep, is inside the \HII\ region
       but close to a dust wall (the right third of the image is peppered with red stars visible only in \JT) at the location of a CO cloud \citep{YangFuku92} and 
       has an intermediate extinction between the extremes of the other three.
 \item The bottom right panel shows a section of Sharpless 2-132. The region contains BD~+55~2722, a system with three O stars and similar extinction values (AB is
       unresolved in our photometry and is analyzed as a single source). A fourth O star, ALS~12\,320 sits at the location of a dust lane that coincides with a 
       CO cloud \citep{Vasqetal10} and has a significantly larger \ebv.
\end{itemize}

\begin{figure}
\centerline{\includegraphics[width=\linewidth]{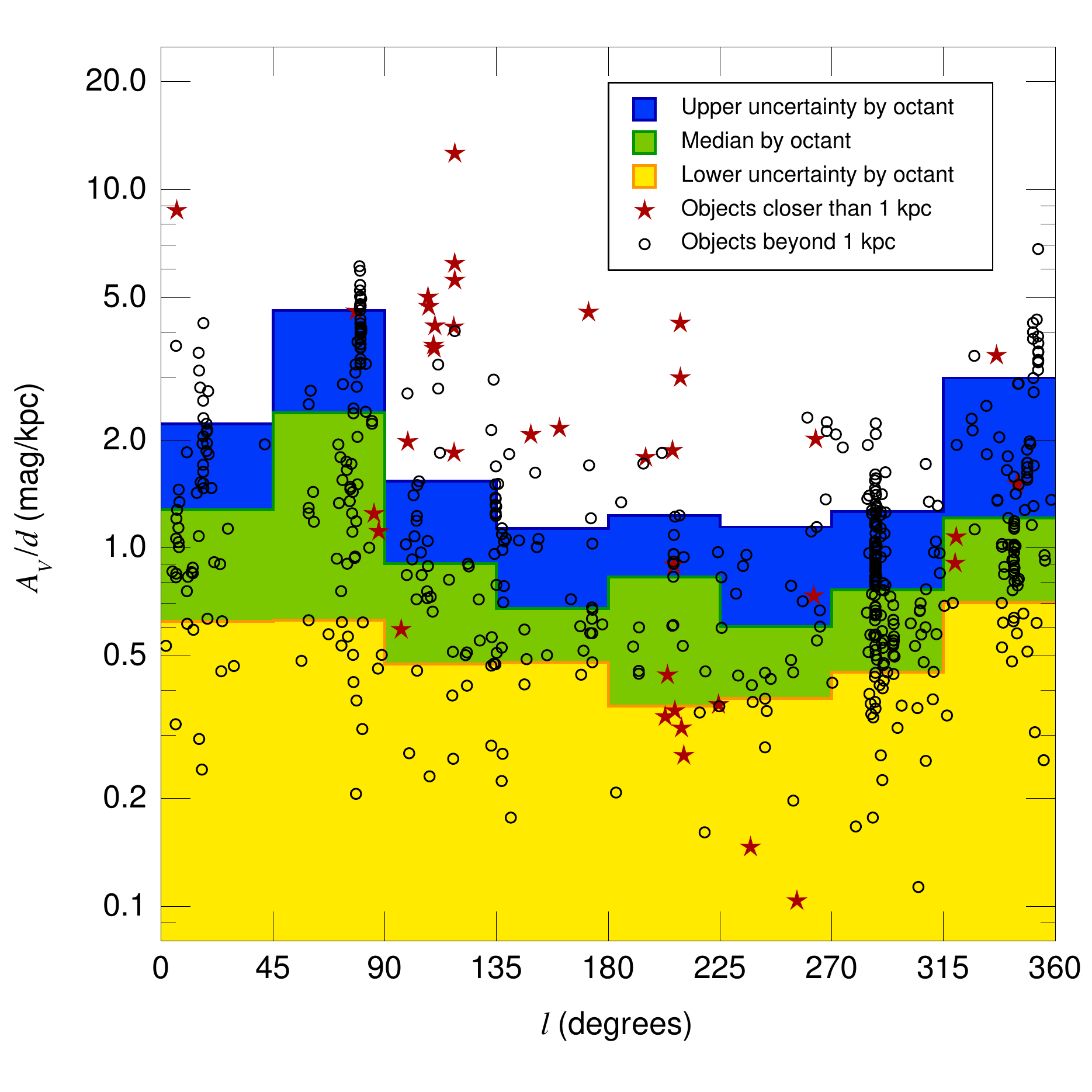}}
\caption{\AV/$d$ for the sample in this paper divided into stars closer than and beyond 1 kpc. No error bars are plotted but they are estimated at $\sim$20\%, 
         with the larger part of the error budget coming from the distance uncertainty. The underlying histograms show the 16th, 50th, and 84th percentiles 
         (median plus lower and upper 1$\sigma$ equivalents) per Galactic octant. We note that only 19 stars have $|b| > 10^{\rm o}$, of which seven are in Orion (see 
         Fig.~\ref{extinction_map}).}
\label{lon_avd}
\end{figure}

These examples show that the explanation behind the differential extinction seen within young stellar clusters and associations lies in the presence of parsec-scale
clouds containing dust and, in many cases, dense enough to be detected in CO. This is, of course, not a new idea, but clearly showing it in these cases is a required 
preliminary step for the relationship with \rv\ that will be analyzed in the next subsection.

In Fig.~\ref{lon_avd} we use the values of \AV\ and \logd\ from Table~\ref{maintable} to plot \AV/$d$, the extinction per unit distance for our sample. We have 
also divided the sample by Galactic octant and calculated the median, and lower and upper 1$\sigma$ equivalents in each one of them. 
Given our understanding of how GOSSS is proceeding, we expect the sample to be complete to $\sim$1 kpc and mostly complete to $\sim$1.5 kpc, though for some
directions incompleteness does not become important until we reach several kpc.
A median value of $\sim$1~mag/kpc is 
a reasonable approximation for the extinction experienced by O stars, in line with previous studies of early-type stars \citep{Fitz68,Necketal80,Forb85}, 
but should be used with care for the following reasons:

\begin{itemize}
 \item There is a considerable dispersion in \AV/$d$ in all of the octants, an indication that some of the extinction is caused by a clumpy medium.
 \item The lower uncertainty, probably a better measurement of the diffuse (non-clumpy) part of extinction, is $\sim$0.5~mag/kpc.
 \item There are differences among octants - the first, second, and eighth (three of the four inner octants) having significantly higher extinctions
       than the other five. 
       As expected, there is more dust towards the inner Milky Way than in the opposite direction.
 \item The two regions that contribute the largest number of objects to our sample, Cygnus in the second octant and Carina in the fifth one, have an effect on the
       histogram. Cygnus is a high-extinction region (e.g., Fig.~\ref{CygXCO}) that produces the largest median and dispersion in any octant 
       and is the direction along which sample incompleteness becomes important at shorter distances.
       Extinction is lower than average towards Carina, in part because many sightlines cross the interarm space between the Sagittarius and Scutum-Centaurus arms.
       Hence, the extinction distribution in the fifth octant resembles the four outer octants, not the other three inner ones
       \footnote{Another consequence of the existence of the interarm space in some fifth-octant sightlines is that we are able to include in our sample stars 
       from NGC~3603, located approximately three times farther away than the Carina Nebula, at a distance similar to that of the Galactic Center but with much lower 
       extinctions.}. 
 \item Stars within 1~kpc of the Sun are concentrated towards the outer octants and show a large scatter than the most distant sample. This is consistent with an ISM
       where extinction is produced by a slowly varying diffuse component and a clumpy component: scatter in \AV/$d$ is expected to increase at shorter distances, where 
       some stars will be dominated by the clumpy component \citep{Lalletal14}. Indeed, the object with the largest value of \AV/$d$ in our sample is Tyc~4026-00424-1 
       (see previous discussion about the bottom left panel in Fig.~\ref{HIIregions1}). 
\end{itemize}

One important characteristic of the solar neighborhood relevant for our analysis is that we are located inside a cavity filled with hot gas called the Local Bubble 
\citep{Snowetal98,Sfeietal99,Maiz01c,Lalletal14}. Typical reddenings measured to stars located at the edge of the Local Bubble ($\sim$~80~pc) at the Galactic plane
are $E(\bS-\yS) = 0.02$ \ \citep{Reisetal11}, with maxima around $E(\bS-\yS) = 0.04$ (which corresponds to $\ebv\sim 0.05$). We note that for \rv~=~3.1 the first of those 
values yields \AV/$d$ of $\sim 1.0$~mag/kpc, which is the typical value measured above for O stars. Therefore, we do not need a dense medium to produce the diffuse
component of Galactic extinction: an ISM as thin as that inside the Local Bubble ($\sim$0.01~atoms~cm$^{-3}$) appears to be sufficient. 

The Local Bubble has a complex shape and, in particular, it has a finger that extends towards the Orion OB1 association \citep{Lalletal14}, which is located at a
distance of $\sim$400~pc and is separated from the Galactic plane (a fact that also contributes to the reduction of material in its sightlines). Some of the BA stars
in the foreground part of the association have near-zero extinction \citep{AlveBouy12}. We confirm this by measuring very small reddenings towards some of the O stars 
in Orion, with \ebv\ values between 0.022 and 0.044 for $\delta$~Ori~AaAb, $\zeta$~Ori~AaAbB, $\sigma$~Ori~AaAbB, $\iota$~Ori, and $\upsilon$~Ori. $\mu$~Col, an O star 
ejected from Orion OB1, has an even lower reddening. The other O-type objects in Orion, $\lambda$~Ori~A, $\theta^1$~Ori~CaCb, and $\theta^2$~Ori~A, have significantly 
higher extinctions due to their surrounding material, something which will we come back to in the next subsection. The irregularity of the ISM around the Local
Bubble is confirmed by the $\zeta$~Oph sightline. That object is the closest O star and is located at a high latitude (it is another runaway star like $\mu$~Col) in a
direction nearly opposite to Orion OB1 but it is the star with the second largest \AV/$d$ in Fig.~\ref{lon_avd}: a good example of the large scatter in the extinction
experienced by nearby objects. We note that the $\zeta$~Oph sightline is used as a reference for the study of the elemental composition of dust in the intervening ISM 
\citep{Drai11}. Also, $\zeta$~Oph is the prototype $\zeta$ sightline for DIBs ($\sigma$~Sco is the prototype $\sigma$ sightline, see
\citealt{Kreletal97,Coxetal05,Maiz15a}), thought to represent an ISM shielded from exposure to UV radiation.

\subsection{The \ebv-\rv\ plane and the properties of dust}

\subsubsection{What our results show}

\begin{figure}
\centerline{\includegraphics[width=\linewidth]{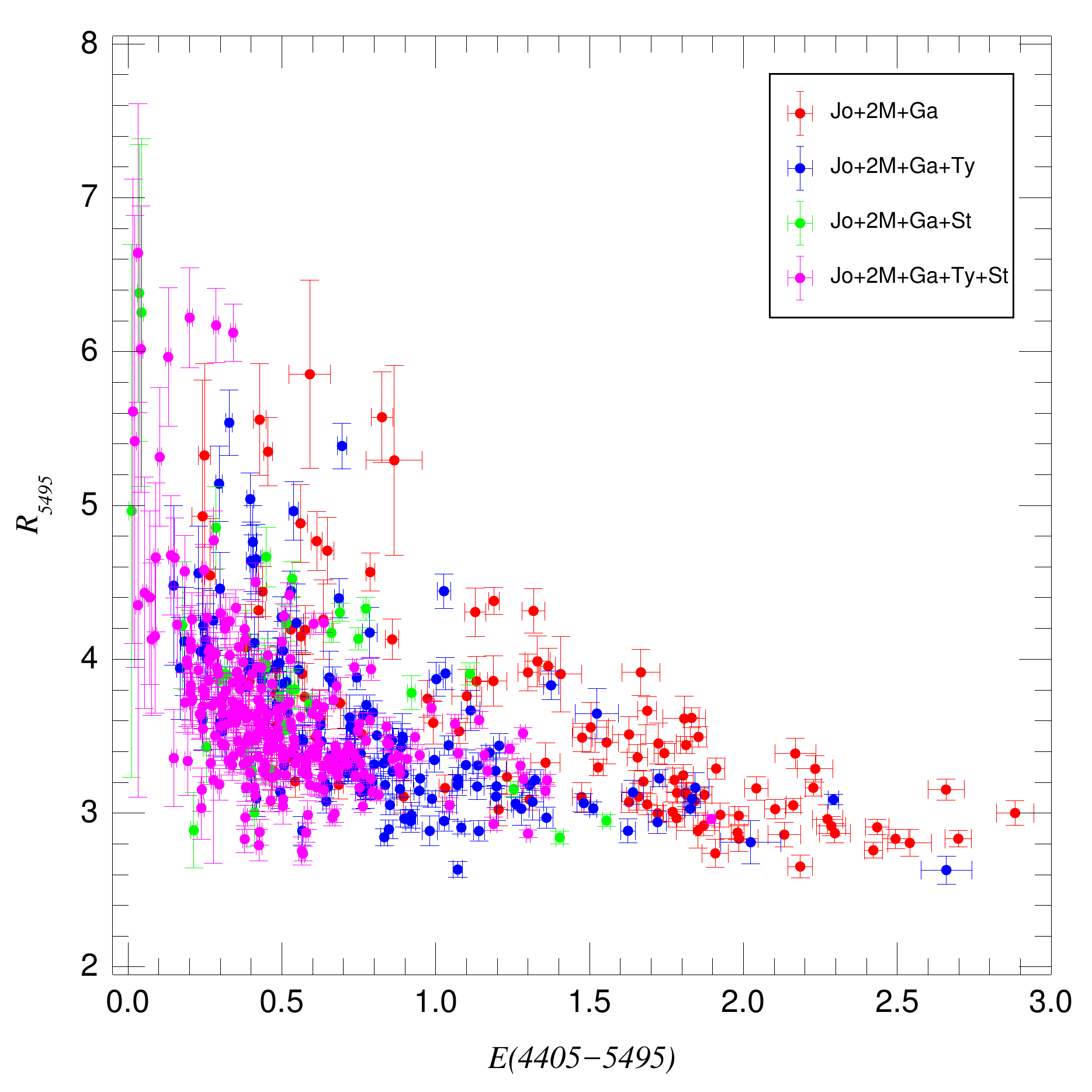}}
\caption{\rv\ as a function of \ebv\ for the CHORIZOS runs in this paper using the MA14 family of extinction laws. The color coding is the same as in 
         Fig.~\ref{ebv_srv}. See Fig.~\ref{rv_histo2} for the histograms derived from this plot by dividing the horizontal axis in five different ranges.}
\label{ebv_rv}
\end{figure}

\begin{figure}
\centerline{\includegraphics[width=\linewidth]{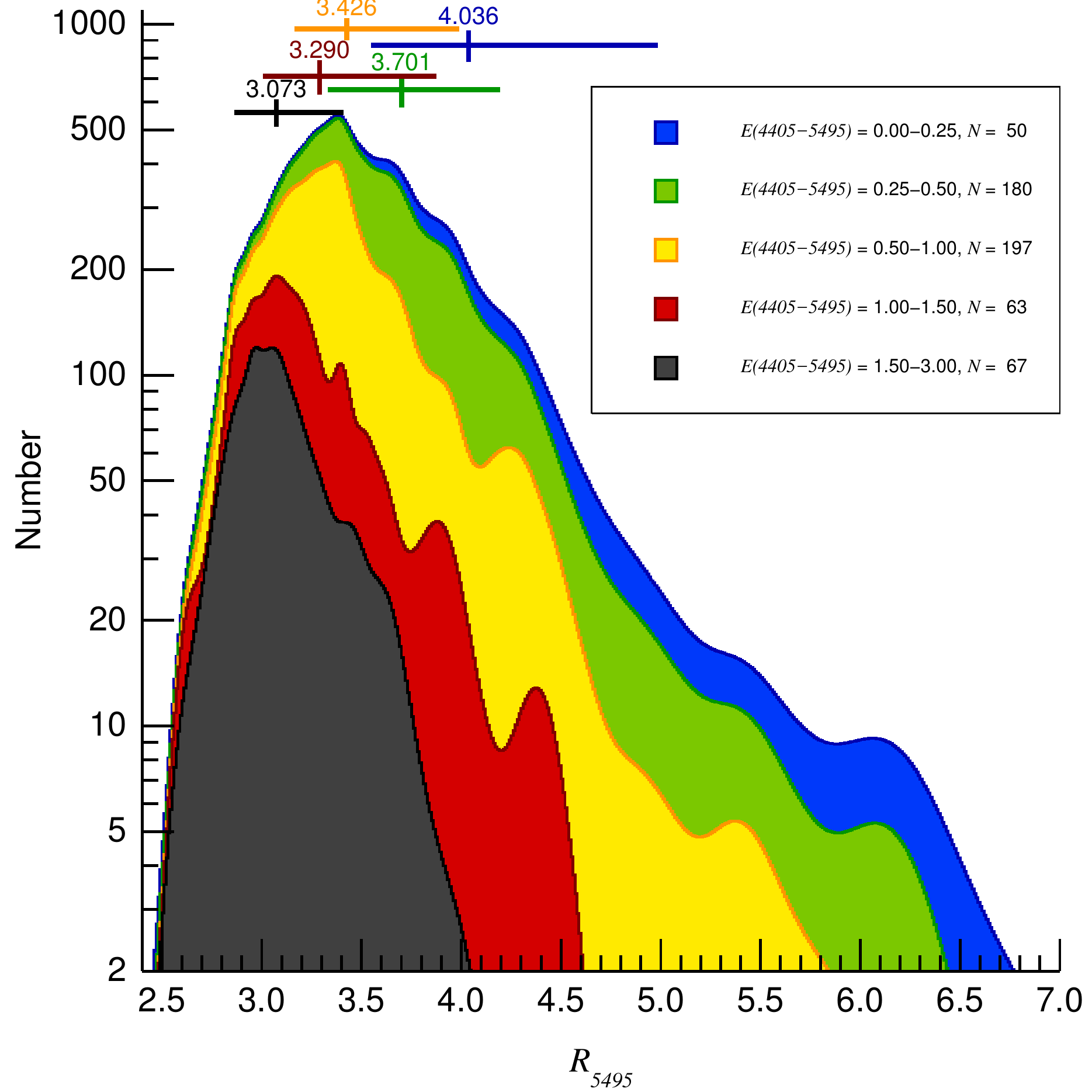}}
\caption{\rv\ cumulative histograms for different values of \ebv\ (see legend for ranges and number of objects in each one) for the CHORIZOS runs in this paper 
         using the MA14 family of extinction laws. Each object is represented by a normalized Gaussian with a width of $\sigma_{R_{5495}}$. We note that the 
         vertical scale is logarithmic. The color crosses near the top mark the median (also printed as the numerical value) and 1$\sigma$ equivalent points for 
         each of the \ebv\ ranges.}
\label{rv_histo2}
\end{figure}

\begin{figure}
\centerline{\includegraphics[width=\linewidth]{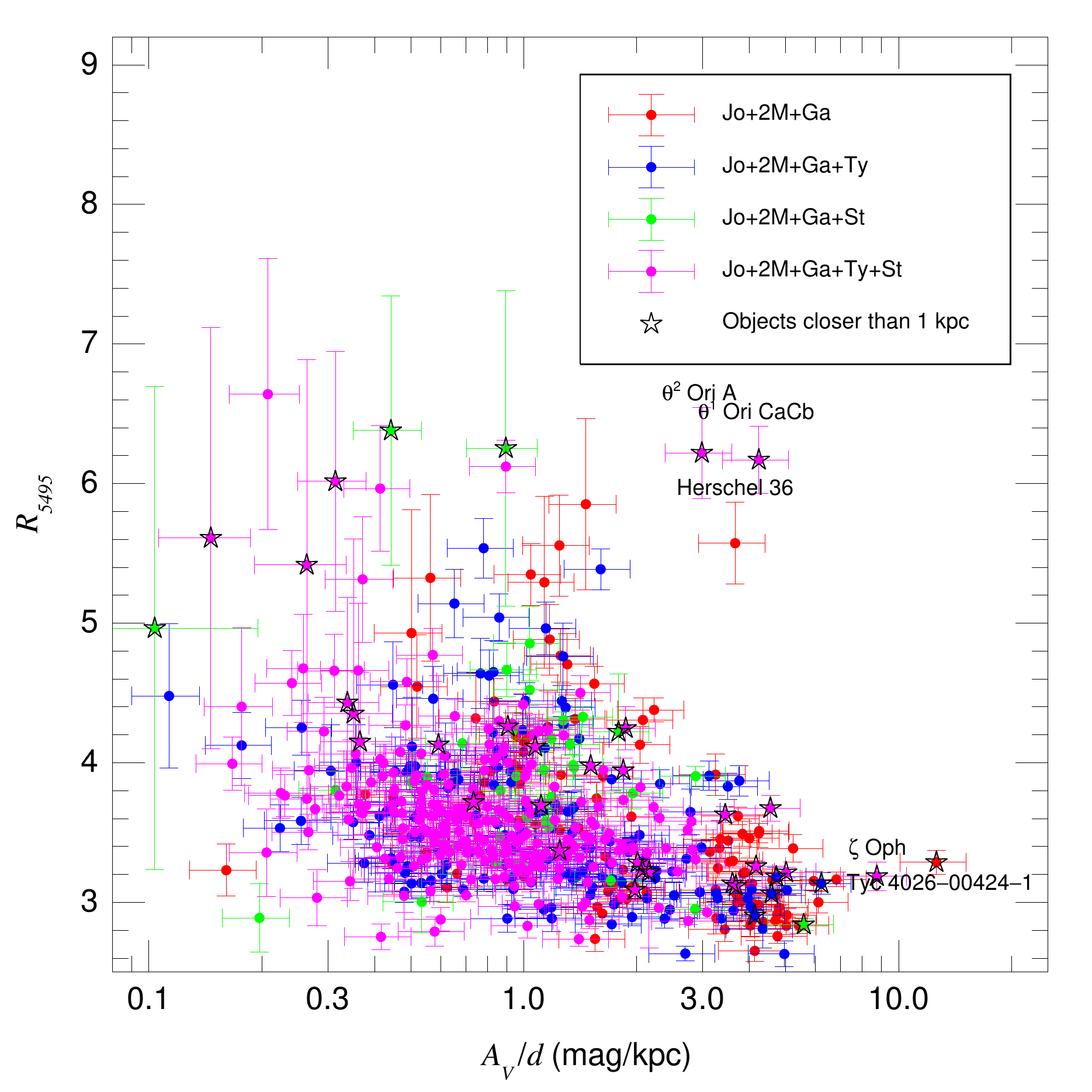}}
\caption{\rv\ as a function of \AV/$d$ for the CHORIZOS runs in this paper using the MA14 family of extinction laws. The color coding is the same as in 
         Fig.~\ref{ebv_srv}. We note that the horizontal scale is logarithmic.}
\label{avd_rv}
\end{figure}

$\,\!$\indent To study the relationship between amount and type of extinction we present three figures and one table. Figure~\ref{ebv_rv} shows the distribution of 
objects in the \ebv-\rv\ plane. Figure~\ref{rv_histo2} shows the \rv\ histograms for five different \ebv\ ranges, from 0.00-0.25 (very low extinction, blue) to 
1.50-3.00 (highest extinction in our sample, dark gray). Figure~\ref{avd_rv} shows \rv\ as a function of \AV/$d$, a measurement of average dust density along the 
sightline (\ebv\ is a measurement of column density). Table~\ref{clusassoc} gives the ranges of \ebv\ and \rv\ for the more relevant clusters and associations.

\input{clusassoc}

The most obvious feature in Figs.~\ref{ebv_rv}~and~\ref{rv_histo2} is the existence of a clear relationship between \ebv\ and \rv. High-reddening stars have low
values of \rv\ (median $\sim$3 for $\ebv > 1.50$) and a narrow distribution. At lower reddenings, both the median and the width of the distribution
are increased. Part of the increase of the width of the distribution is due to the increase in the individual values of $\sigma_{R_{5495}}$ (Fig.~\ref{ebv_srv}) but that is 
a minor effect: at high reddenings there are no examples with large values of \rv\ while for low reddenings there are cases with both high and low values of \rv\ with
low uncertainties. When we reach the $\ebv < 0.25$ range, the median \rv\ value is $\sim$4. Figure~\ref{avd_rv} is relatively similar to Fig.~\ref{ebv_rv} but five
objects are conspicuously placed: the previously discussed Tyc~4026-00424-1 and $\zeta$~Oph at low \rv\ values and three more at high \rv\ values, Herschel 36, 
$\theta^2$~Ori~A, and $\theta^1$~Ori~CaCb. We analyze these later in this paper.

In the next subsubsection we combine these results with those of other authors to build a picture of how \rv\ varies with ISM phases. Before doing that, we address a
preliminary point: when a sightline has two or more clouds of different \rv\ and extinction is described by the MA14 (or CCM) families, the resulting extinction law
is also a member of the family with the equivalent \rv\ being an average of the individual values weighted by the individul reddenings (see Appendix C of MA14). This can
be seen as an advantage or an inconvenience: it is the first from the point of view of correcting for extinction because it provides a universality to extinction laws 
but it is the second from the point of view of studying dust because it partially erases the information contained in the individual clouds.

Some papers (e.g., \citealt{Valeetal03}) use the above property to measure extinction decomposing between a foreground and a cluster component. The foreground component has 
\rv~=~3.1 and the \rv\ of the cluster extinction is measured by the paper. We have decided not to do that in this paper for the following three reasons: 

\begin{itemize}
 \item While \rv\ = 3.1 may be close to the average Galactic extinction, it is not necessarily the case for every sightline. For example, some objects with low 
       extinction have large values of \rv, so the foreground (likely the only extinction) cannot have \rv\ = 3.1. 
 \item The decomposition between two components usually assumes that the foreground extinction is constant in space but that may also be false. See the 
       interesting case of 30 Doradus \citep{vanLetal13}, where the velocity separation between the Galactic (``foreground'') and LMC (``cluster'') ISM lines allows for 
       maps of both components. See also \citet{Belletal17} for a detailed study of the variations in the foreground extinction to a globular cluster.
 \item The assumption of a constant \rv\ for the cluster component is wrong in some cases (see Fig~\ref{HIIregions1} and discussion below). 
\end{itemize}

\subsubsection{Consistency with other results: \rv\ of different ISM phases}



$\,\!$\indent There are previous studies of the distribution of \rv\ in the Galaxy, but most of them refer to single clusters or associations, not to samples
that cover the whole Galactic plane or a significant part of it. We start by looking at the ones that refer to objects listed in Table~\ref{clusassoc}: 
\citet{Belietal99} find a variable $R_V$ between 2.8 and 3.9 in M16, \citet{Pangetal16} a variable $R_V$ between 2.48 and 4.06 in NGC~3603, \citet{Limetal14} a 
canonical extinction law (\rv = 3.1) towards NGC~1893, and \citet{Vazqetal96} $R_V = 4.70\pm0.65$ in the Carina Nebula. Going to older 
studies, the high value of \rv\ in the Orion Nebula has been known for a long time, as it has been the prototype for a non-canonical extinction law since the 1930s
\citep{BaadMink37}. All of these results are consistent with ours, with some small differences that can be explained by the sample and method differences. 

In a second group, we look at modern, large-area studies of extinction type. \citet{Schletal16}, a paper we already discussed in the previous subsection, obtain
$R_V$ results different from ours: they find an average value of 3.32, a dispersion of 0.18, and no significant variation with the amount of extinction. There
are no tails in their distribution, so few stars have values under 3.0 and even less values over 4.0. Another study is that of \citet{FitzMass07}, who find a mean 
value for $R_V$ of 2.99 and a dispersion of 0.27, with a long tail that extends to large values beyond 4.0. How can we reconcile the different results from the three
papers? We consider the following points in the discusion below:

\begin{itemize}
 \item \citet{FitzMass07} do not distinguish between \rv\ and $R_V$. They claim to give $R_V$ but we suspect it is \rv\ they are working with. However, their 
       sample consists of OB stars with typical values of $E(B-V)$ around 0.45, so the differences between the two should be small (see Fig.~3 in \citealt{Maiz13b}). 
       A similar argument can be made about $E(B-V)$ and \ebv. \citet{Schletal16} use the approximation of defining monochromatic wavelengths for each filter, 
       therefore ignoring non-linear effects in the reddening trajectories. This could bias their measurements when comparing high and low extinctions.
 \item \citet{FitzMass07} use \UJ\BJ\VJ\JT\HT\KT\ photometry, so their wavelength coverage is similar to ours. \citet{Schletal16}, on the other hand, use a 
       filter set weighted towards longer wavelengths.
 \item Most importantly, the three samples are different in terms of targets, amount of extinction, and average distances. \citet{FitzMass07} sample is
       relatively similar to ours but it is biased towards B stars, which tend to be older and to be located farther away from \HII\ regions. The amount of extinction 
       range it covers is also smaller (only low values) and their sample is less concentrated towards the Galactic plane and is located at shorter distances (on
       average) compared to ours. \citet{Schletal16} sample is fundamentally different from the other two, as it is much larger and consists of red giants and red 
       clump stars, with very few of them (if any) located close to \HII\ regions. The distribution of the amount of extinction is similar to ours but they lack a significant 
       fraction of close objects, as their typical distances are between 1 and 5 kpc.
\end{itemize}

Another example of the second group of studies is \citet{Heetal95}, who analyzed the extinction experienced by a sample of southern OB stars with $\ebv < 2$. \cite{Heetal95}
is affected by some of the issues discussed in the appendices in this paper and it also has the additional problem that most of its ``spectral types'' are actually 
photometric classifications (i.e., not true spectral types), which are known to have frequent large errors. Furthermore, many of their true spectral types are obtained 
from the Michigan Catalog (whose last volume was published as \citealt{HoukSwif99}), which also contains many errors due to the imprecisions associated with objective-prism
spectroscopy (poor spectral resolution, source confusion, and nebular contamination), something that can be checked by comparing the spectrograms published in GOSSS I+II+III 
with the Michigan spectral types. As an example of the errors, ALS~4923 is listed in \citet{Heetal95} as having a spectral type of O6~V as derived from photometry and of 
O9.5~III as derived from spectroscopy \citep{VijaDril93}. The GOSSS spectral type shows that it is actually an O8.5~V~+~O8.5~V SB2 (see GOSSS III for a spectrogram). 
Therefore, \citet{Heetal95} is likely to include systematic errors and an analysis of the sample in common with the stars in this paper shows that it tends to underestimate 
\AV. Nevertheless, the paper has an interesting result that is consistent with what we find here: their measured average $R_V$ decreases with distance, being 3.31 for stars 
closer than 1~kpc and 2.98 for stars farther away than 5~kpc (see Fig.~\ref{rv_histo2} - we note that more distant stars are, on average, more extinguished than closer 
ones). 

\begin{figure*}
\centerline{\includegraphics[width=\linewidth]{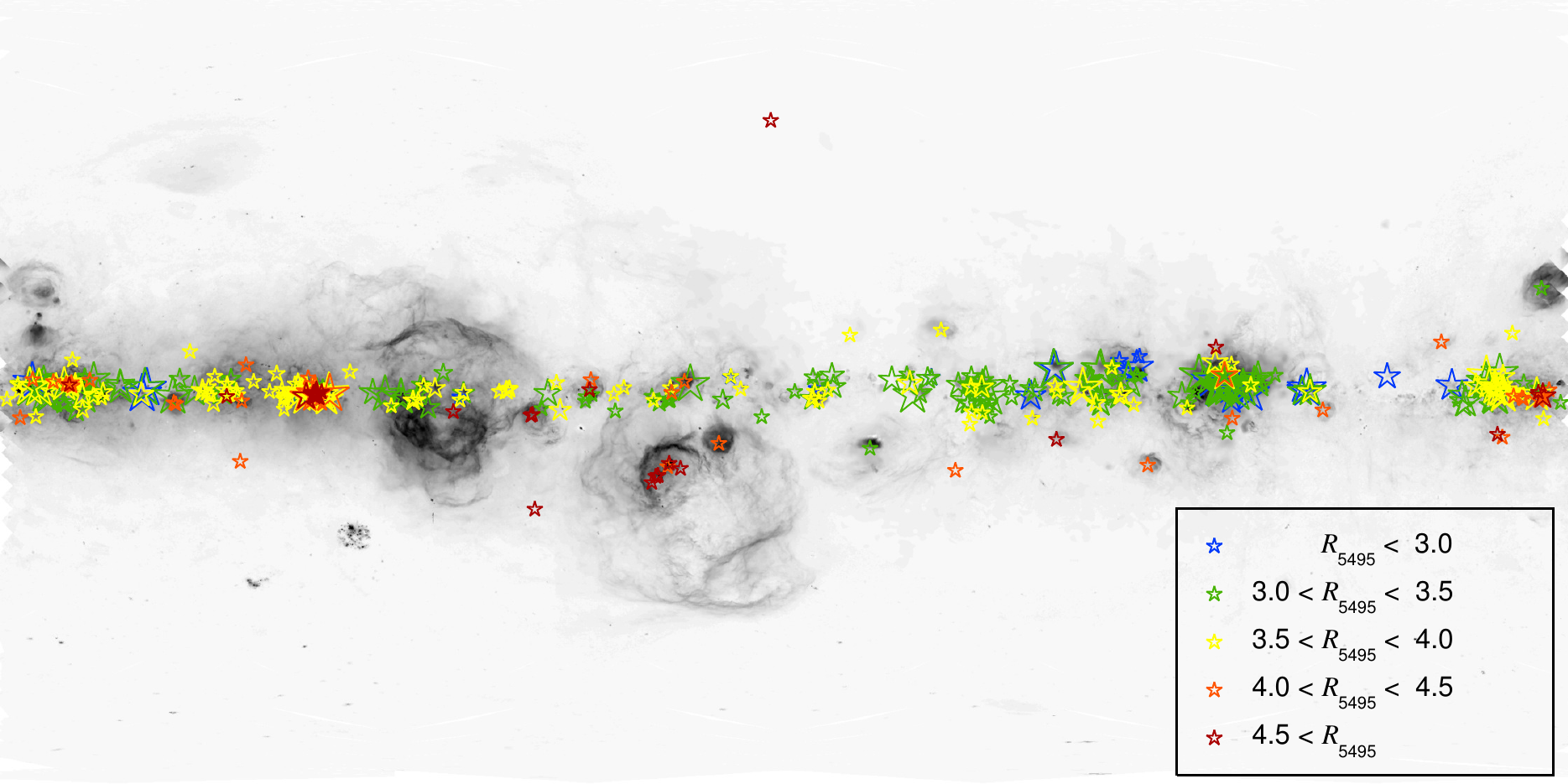}}
\caption{Extinction measurements for Galactic O stars plotted on top of the full-sky H$\alpha$ image of \citet{Fink03}. Different colors are used for five ranges of \rv\ with
         the size of the symbol increasing with \ebv. The background image is showed in a logarithmic intensity scale and in Galactic coordinates with a Cartesian projection centered 
         on the Galactic anticenter.}
\label{extinction_map}
\end{figure*}

In a third group, we look at studies of individual clusters and associations not included in this paper. In the case of \HII\ regions, Westerlund~2 
\citep{Zeidetal15,Huretal15}, NGC~1931 \citep{Limetal15}, and 30 Doradus (MA14) have values of $\rv > 4$, in the case of 30 Doradus with a large scatter (see below).
For objects without an \HII\ region, \citet{Straetal14} find an $R_V$ of 2.87$\pm$0.16 for NGC 6913, a cluster in Cygnus with an intermediate extinction, and
\citet{Marcetal14} find eight stars in VdBH 222 with \rv\ between 2.7 and 3.0 and a high reddening (\ebv\ between 2.5 and 2.9).

The results in this paper and in the previously listed references can be explained if we model the dust in the ISM within several kpc of the Sun as having three different regimes 
depending on the ambient UV radiation level:

\begin{itemize}
 \item {\bf Regions with high levels of UV radiation have large values of \rv\ ($>4$).} This includes \HII\ regions but also cavities filled with hot gas such as the Local 
       Bubble where UV radiation can travel relatively unimpeded. \HII\ regions can have significant dust densities but, given their small sizes, only in exceptional 
       situations they provide column densities large enough to produce strong extinctions. Cavities occupy much larger volumes but they have low dust 
       densities so their signature is easily erased when the sightline includes denser clouds.
 \item {\bf Regions with low levels of UV radiation have small values of \rv\ ($<3$).} These regions are those with significant column densities of CO or, alternatively,
       those that are easily detected as dust lanes in the optical. They occupy a small fraction of the ISM volume but they have the highest dust densities, so their 
       \rv-reducing signature is harder to erase than that of cavities with high values of \rv\footnote{Further than a few kpc from the Sun it is possible that similar 
       conditions are attained in other environments due to the dearth of sources of ionizing radiation but such locations are not covered by the sample in this paper.}.
 \item {\bf Regions with intermediate levels of UV radiation have intermediate values of \rv\ (between 3 and 4).} These regions represent a typical warm to 
       cold ISM (excluding \HII\ regions, cavities, and molecular clouds) that fills the majority of the volume in the Galactic disk. 
\end{itemize}

That model can explain all of the results previously described. OB stars in the Local Bubble (and adjacent bubbles) detected in the samples in this paper and in the
\citet{FitzMass07} sample will have low values of \ebv\ and high values of \rv\ with a relatively large scatter caused by large uncertainties and the presence of small clouds
that provide partial shielding from UV radiation. This is what is observed for the off-plane stars in Fig.~\ref{extinction_map}, which are relatively nearby and for which the Local
Bubble contribution to extinction should be generally larger than for objects closer to the Galactic plane.
We note that no \citet{Schletal16} objects are present in the Local Bubble. Moving to longer distances (and ignoring \HII\ regions for the moment) we see
mostly the effect of the typical ISM in the three samples, as the signature of the Local Bubble is easily erased. This region (up to 1-2~kpc) is the classical regime for 
extinction studies and the origin of the definition of 3.1 as the canonical \rv\ value, although some of the objects here (e.g., those in Cyg OB2) already show the effect
of molecular-cloud extinction. Once we start moving out, the \citet{FitzMass07} sample disappears and we are left only with the other two samples. Cool stars will likely
not be associated with molecular clouds, so their \rv\ should not change much, especially if they are in one of the outer Galactic quadrants 
(we note that the \citet{Schletal16} does not
include objects in the fourth quadrant). O stars in the inner quadrants (excepting those whose sightlines are dominated by an interarm space such as those in NGC~3603)
will likely have a molecular cloud in their sightline and will end up having low values of \rv. 

\begin{figure*}
\centerline{\includegraphics[width=0.49\linewidth]{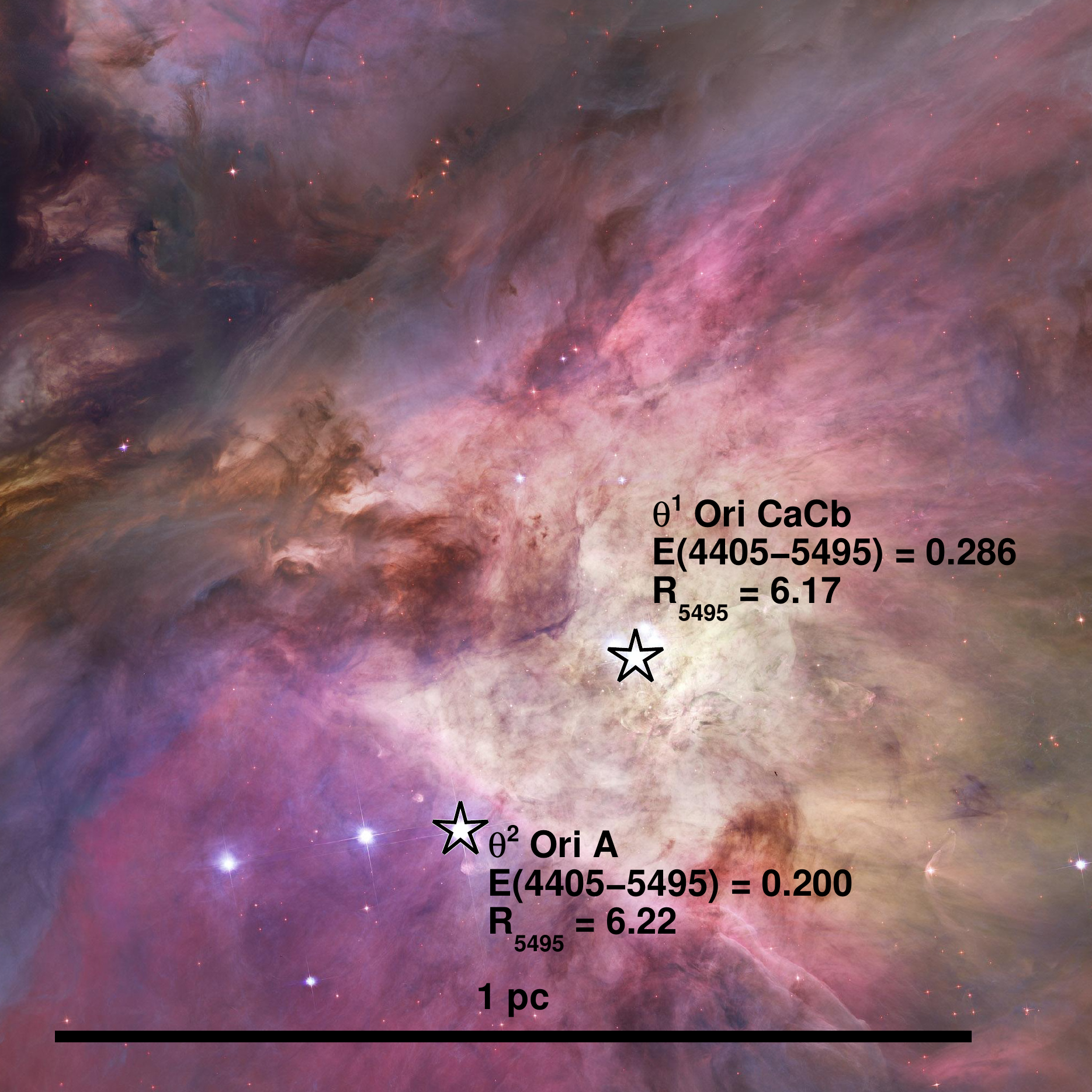} \
            \includegraphics[width=0.49\linewidth]{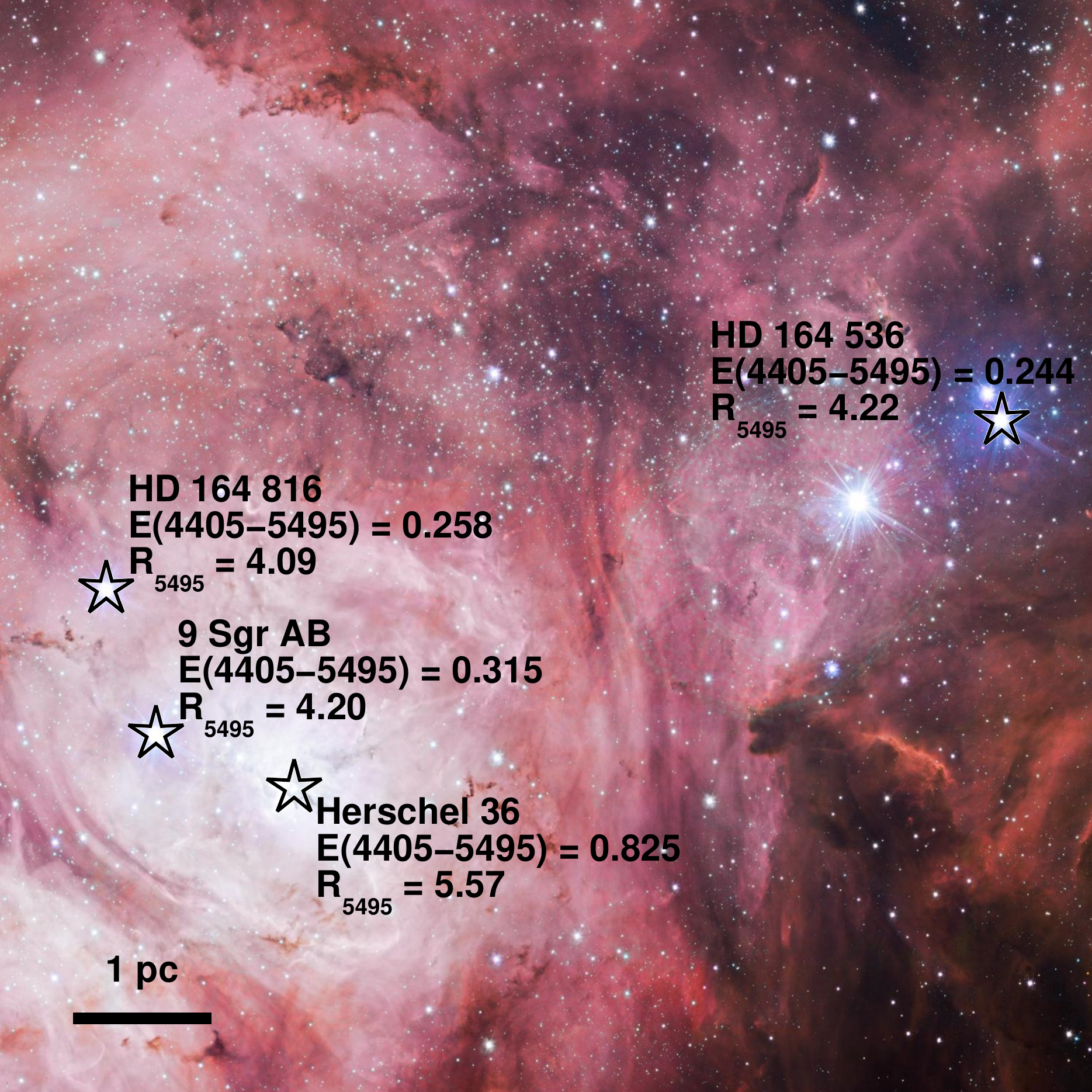}}
 \caption{Extinction measurements for O stars in two \HII\ regions: (left) the Orion Nebula, (right) M8. The left image is from STScI press release 2006-01 
          ($BV$H$\alpha iz$) and the right image is from ESO press release 1403 ($ugr$H$\alpha i$). North is towards the top and east to the left. The approximate 
          physical scale is indicated in both cases.}
\label{HIIregions2}
\end{figure*}

As we have seen in the previous subsection, \HII\ regions are complicated in terms of extinction because the sightlines towards different stars cross different environments:
the \HII\ region itself (first type), possibly a molecular cloud (third type), and in most cases the diffuse, typical ISM (second type). Therefore, almost every value
of \rv\ is possible. In three of the panels of Fig.~\ref{HIIregions1} there is a clear anticorrelation between \ebv\ and \rv: {\bf when we approach a dust lane \ebv\ increases,
as we had already discussed, but \rv\ decreases.} This is consistent with our model, where \rv\ is correlated with UV flux. The fourth panel, NGC 7822/Berkeley 59, shows a
different behavior with the four sightlines having $\rv < 3$. The likely explanation is that most of the extinction common to the four sightlines is coming from a molecular 
cloud that affects all sightlines, which is consistent with the four stars having large values of \AV/$d$. To understand \HII\ region extinction better, we should look at three 
of the extreme cases in Fig.~\ref{avd_rv}: $\theta^1$~Ori~CaCb and $\theta^2$~Ori~A, the two O stars in the Orion Nebula, and Herschel 36 in M8 (Fig.~\ref{HIIregions2}):

\begin{itemize}
 \item As we have previously mentioned, the foreground extinction towards the Orion OB1 association (including the Orion Nebula) is very small. The reddenings of the two O 
       stars in the Orion Nebula are significantly higher, indicating that their extinction must be fundamentally local. Evidence for this effect has been found by 
       \citet{vadWetal13}, who estimate that the extinction for Trapezium stars takes place within 1-2 pc, and by \citet{ODelHarr10}, who determine that the nebular 
       extinction at the edges of the nebula essentially goes to zero. $\theta^1$~Ori~CaCb is closer to the center of the nebula than $\theta^2$~Ori~A, as it is the main
       ionization source, and correspondingly has a higher extinction. Those circumstances explain why the Orion Nebula stars are the prototype for high \rv\ extinction: they
       are located in the (nebular) bright part of an \HII\ region with very little intervening material in the sightline (either as nearby molecular clouds or as a typical
       ISM). That configuration yields $\rv > 6$.
 \item The foreground extinction towards M8 is lower than average for its distance, as evidenced by the low \ebv\ of HD~164\,536 and HD~164\,816, and likely caused by the
       in-between region being an interarm space. Looking at the right panel in Fig.~\ref{HIIregions2} we see that Herschel 36 has both significantly larger \ebv\ and \rv.
       Compare this to the opposite relationship between those two quantities in, for example, the Carina Nebula. The explanation is the geometry described by \citet{Maizetal15d}: 
       Herschel 36 is seen through a tunnel carved in the molecular cloud where the gas is exposed to the strong ionizing radiation of the Herschel 36 multiple O-star 
       system \citep{Ariaetal10,SanBetal14}\footnote{Herschel~36 has a nearby companion that dominates the flux in the MIR \citep{Gotoetal06} and which is already 
       significant in the NIR. The existence of that companion complicates extinction measurements and produces discrepancies between different works depending on the
       photometry, extinction laws, and techniques used (see \citealt{Ariaetal06} and \citealt{Maizetal15d}). We note, however, that all results consistently yield \ebv\ 
       values in the 
       0.78-0.85 range and \rv\ values in the 5.4-5.9 range.}.
       The resulting long optical depth yields not only a large \rv\ but also a large \ebv.
\end{itemize}

The model presented here implies that the Herschel 36 case (simultaneous high \ebv\ and high \rv) is difficult to find. Outside of the sample in this paper, only in \citet{Damietal17a}
we find some more extreme examples: four stars with $\ebv > 1.3$ and $\rv > 4.0$ and three stars with $\ebv > 2.0$ and $\rv > 3.5$. Those stars would be placed in an empty 
region in Fig.~\ref{ebv_rv} and they are all in the Carina Nebula. That is a logical place to find such objects, as the foreground extinction is low (the sightline is mostly an
interarm region) and the Carina Nebula is the brightest and largest \HII\ region in the solar neighborhood, thus making it possible to have large column densities of material exposed
to UV radiation.

\begin{figure*}
\centerline{\includegraphics[width=0.33\linewidth]{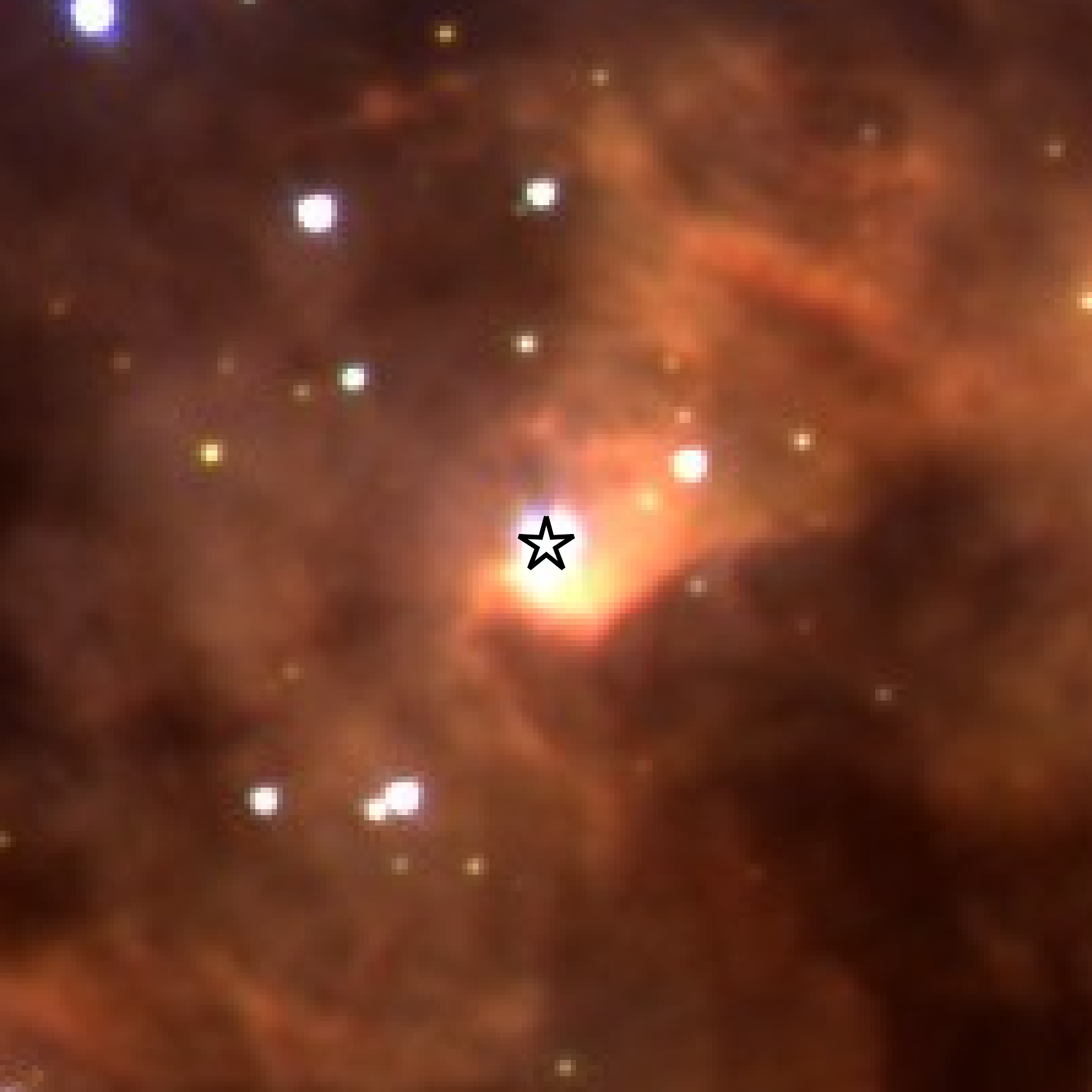} \
            \includegraphics[width=0.33\linewidth]{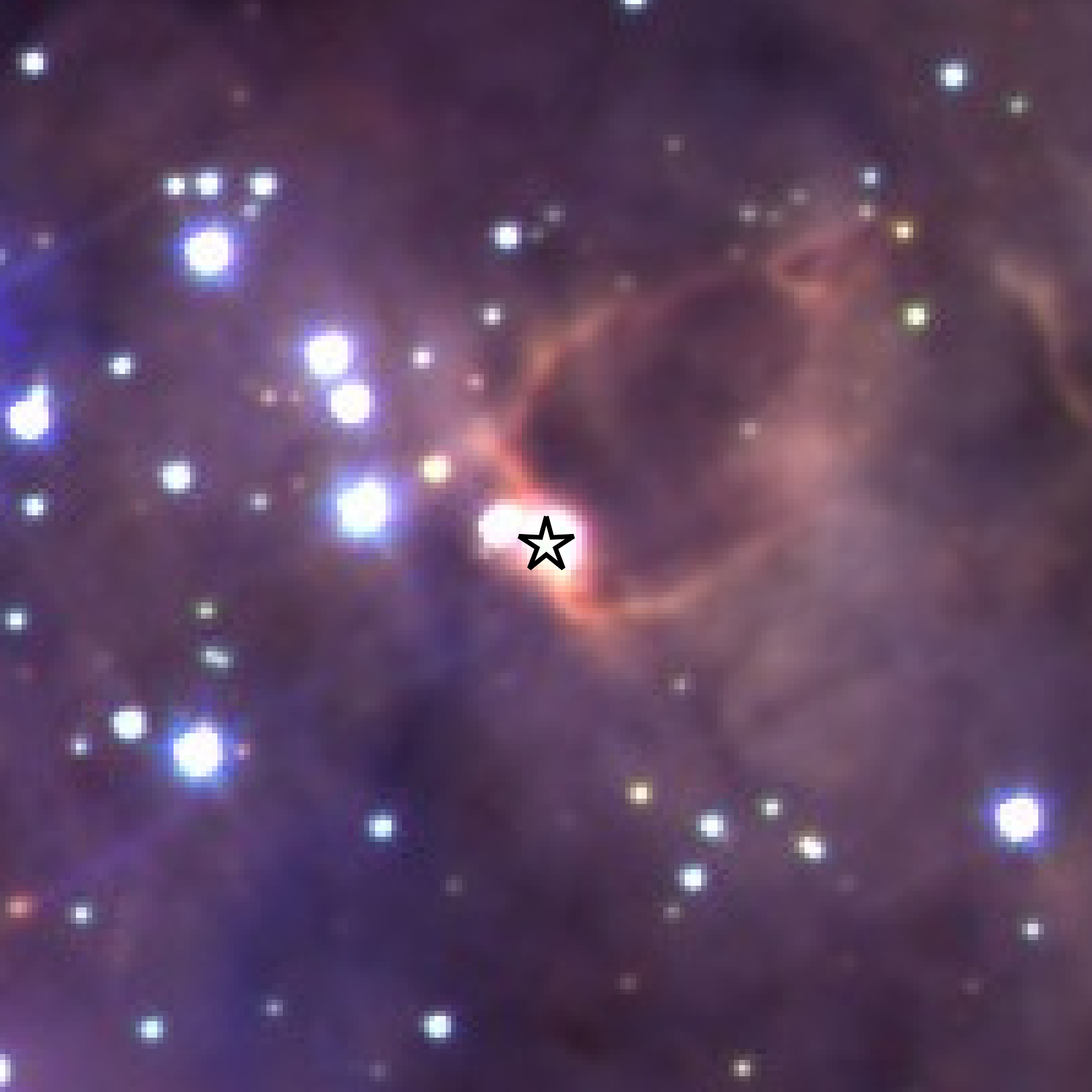} \
            \includegraphics[width=0.33\linewidth]{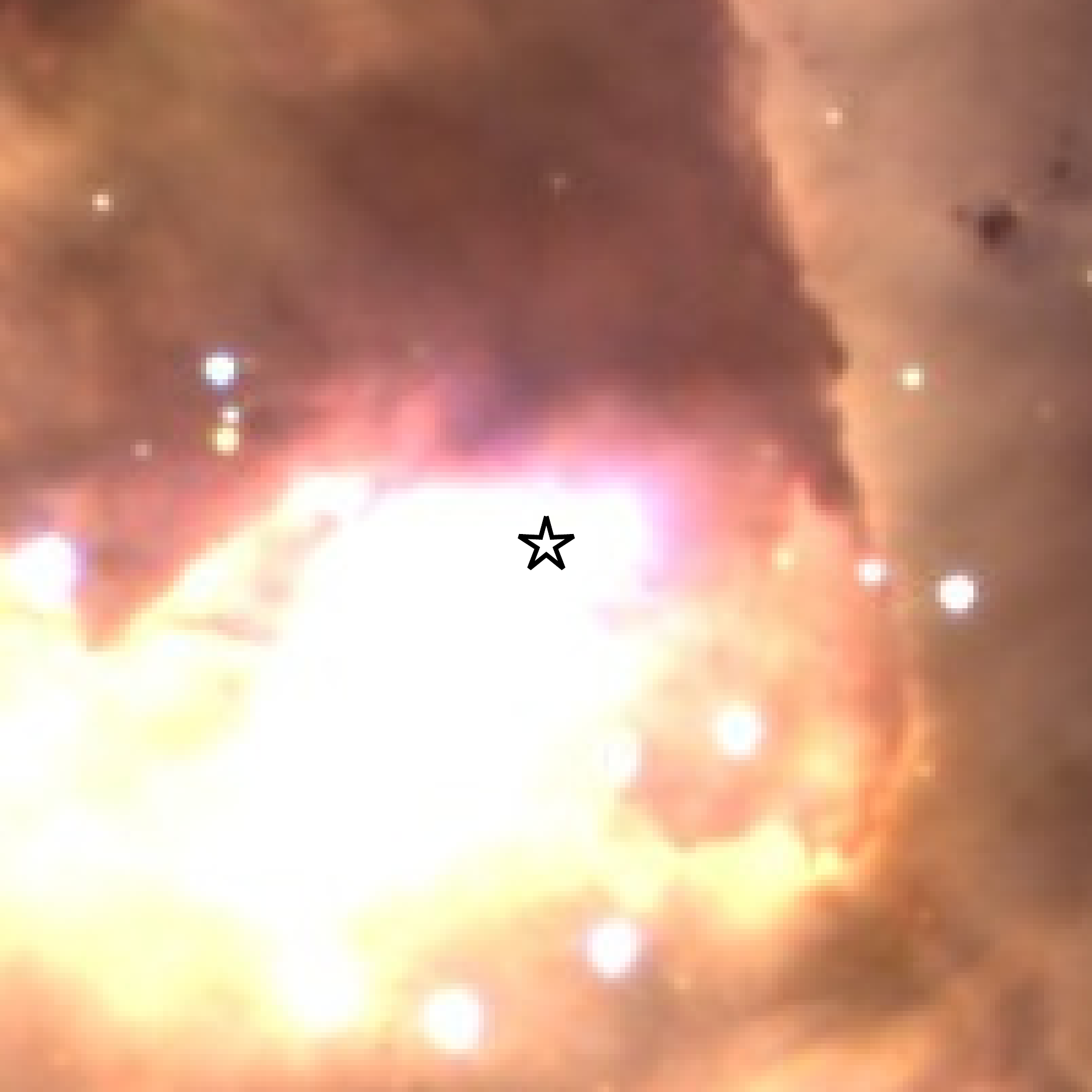}}
 \caption{Environment around VFTS~451 (left), VFTS~464 = [P93]~702 (center), and VFTS~579 = [P93]~1201 (right), the three stars from the MA14 sample in 30 Doradus with 
          $\ebv > 0.3$ and $\rv > 6.0$. The three objects are at the top of pillars that point towards R136.
          The background image is a F814W + F656N (red) + F555W (green) + F336W (blue) WFC3 mosaic built from WFC3-ERS data. Each field is $8\arcsec\times 8\arcsec$ 
          (2 pc $\times$ 2 pc) with north to the top and east to the left. See \citet{Walbetal02a} for an alternative view of knot 1, the environment around VFTS~579.}
\label{VFTS}
\end{figure*}

A final check on our model can be performed with the MA14 extinction analysis of 30 Doradus, where the foreground extinction is low and a wide range of values of \rv\ ca be found 
more easily than for most Galactic \HII\ regions. We have selected the three stars with $\ebv > 0.3$ and $\rv > 6.0$ in the sample, 
VFTS~451, VFTS~464, and VFTS~579 (Fig.~\ref{VFTS}), that is, the three stars with more \HII-like extinction. Where are they located in 30 Doradus? They are in different positions but
in the same type of environment: a compact \HII\ region at the top of a dust pillar created by the radiation and winds from R136. The first two are in relatively small pillars while 
VFTS~579 is in a large one called knot~1 by \citet{WalbBlad87}. Therefore, their environment is precisely the one predicted by our model, as they are immersed in a dense gas 
subjected to a strong UV field.

\subsubsection{The physics behind \rv\ variations}

$\,\!$\indent \citet{Card88} found anticorrelations between \rv\ and either log(H$_2$/\AV) or log(CH/\AV), where H$_2$ and CH represent the column densities of those molecules.
He also found similar results when substituting \AV\ by the total hydrogen column density. \citet{Card88} concluded that the decrease in the abundance of H$_2$ when \rv\ increases was a
combination of (a) a reduction in the formation rate of H$_2$ due to the smaller total grain surface per unit mass for larger grains and (b) an increase in photodestruction via a
decrease in UV dust extinction, as \rv\ anticorrelates with $A_{\rm UV}$/\AV\ (CCM). 

Our results are similar to
\citet{Card88} in the sense that we find that stars behind CO clouds or dust lanes have low values of \rv. However, an important difference is that
his sample is biased towards low extinctions (only two stars have \AV$>$3) and small grain sizes (only seven stars in the 4.0$\le$\rv$<$4.4 range and
none above that). In other words, \citet{Card88} does not include objects in \HII\ regions or subject to Local-Bubble-only extinction, where the physical conditions are different than the 
more stable and benign considered by him. \HII\ regions are short-lived dynamic structures where destruction processes dominate formation ones and regions inside bubbles are too thin for 
dust grains to form. Therefore, we propose that {\bf the ISM regions with large values of \rv\ are produced by the selective destruction of smaller dust grains with respect to large grains.}

What processes could be responsible for such selective destruction? One possibility is thermal sputtering, the erosion of grains by impacting thermal atoms or ions \citep{Drai11}. The grain
lifetime in such a high-temperature is proportional to its size, thus reproducing the observed behavior. Thermal sputtering is insignificant around $10^4$~K but can become important above 
$10^5$~K. That makes it an unlikely source for grain destruction in \HII\ regions but a possible candidate in regions like the Local Bubble.
An alternative destruction mechanism is heating by EUV radiation \citep{GuhaDrai89,Jone04}, which also acts preferentially on small grains. This is the likely cause in \HII\ regions, where
the sources of EUV photons are the O stars.

We point out that dust extinction is not the only ISM observable that is affected by the strength of the UV field. Another observable is the ratio of the equivalent widths of 
two diffuse interstellar bands (or DIBs), DIB~5797 and DIB~5780 \citep{Camietal97,Vosetal11,Maiz15a,Maizetal15c}. $W(5797)/W(5780)$ can be as large as $\sim$0.7 or smaller than 0.1. Sightlines 
with large ratios are produced by $\zeta$ clouds (after the prototype sightline of $\zeta$~Oph) while small ratios are produced by $\sigma$ clouds (after the prototype sightline of 
$\sigma$~Sco)\footnote{\citet{Camietal97} refer to the extreme $\sigma$ sightlines as of Orion type, since stars in the Orion Nebula have a very weak or non existent DIB~5797.}. Even though
the carriers themselves are not identified at this point, the observed relationship indicates that DIB~5797 is found only in regions shielded from UV radiation while DIB~5780 can originate
in either shielded or unshielded regions. This led \citet{Camietal97} to hypothesize a skin effect in ISM clouds, where the core ($\zeta$ sightlines) includes the carriers of both DIBs and 
the outer layers ($\sigma$ sightlines) only the DIB~5780 carrier. The correlation between $W(5797)/W(5780)$ and \rv\ was detected by \citet{Camietal97} and confirmed in a much larger sample by
\citet{Maiz15a}. 


%
%

\section{Conclusions and future work}

$\,\!$\indent The main conclusions of this paper are:

\begin{itemize}
 \item The MA14 family of extinction laws provide a better description than the CCM or F99 ones for the conditions described in this paper: Galactic O stars with optical-NIR photometry up to
       \ebv = 3 (the response to the question we ask in the introduction). However, there are signs that in other cases such as higher reddenings or IR-only data MA14 fails (as expected), 
       so further work is required to generate an improved new family.
 \item Many O stars have extinctions similar to those of nearby late-type objects but some are located close to obscuring material that increase their extinction and can change the effective
       value of \rv.
 \item Young stellar clusters and associations can have large variations in the amount and type of extinction from star to star. In those cases the use of average
       \ebv\ and \rv\ values fails and one needs to determine the amount and type of extinction star by star. Some notorious examples include \HII\ regions such as the Carina Nebula and M8.
 \item The average dust grain size and, hence, \rv, is determined by the level of UV radiation. On the one extreme, shielded ISM regions such as molecular clouds have small grains (low \rv)
       and exposed ISM regions such as \HII\ regions have large grains (high \rv), with the average diffuse ISM in an intermediate position in size and \rv. Several mechanisms have been 
       proposed to explain this relationship but we expect that the dominant one for the existence of high \rv\ regions is the selective destruction of small grains by heating by EUV radiation.
\end{itemize}

Our lines of future work include the following:

\begin{itemize}
 \item We plan to analyze the extinction law that affects OB stars in the IR using a combination of 2MASS, Spitzer, WISE, and ISO data for stars for which we also have GOSSS optical
       spectroscopy. See \citet{Maizetal17b} for some examples.
 \item For the extinction-law families described in this paper the optical part was derived from photometric data and later interpolated in wavelength. An alternative method that yields more
       information and potentially eliminates imprecisions is to use spectrophotometry. To that purpose, we have obtained HST/STIS 1700-10\,200~\AA\ data for several tens of stars in 
       30~Doradus (GO program 14\,104) which we are currently analyzing.
 \item There are two opposing views on the correlation between UV and IR extinction. On the one hand, CCM claim that they are tightly correlated in the sense that large-\rv\ optical-NIR
       extinctions correspond to flat UV extinctions while small-\rv\ optical-NIR extinctions correspond to steep UV extinctions. On the other hand, \citet{FitzMass07} claim that with the 
       exception of some extreme cases, the UV and IR portions of Galactic extinction curves are not correlated with each other. We plan to use the results of this paper in combination with 
       IUE data to analyze those opposing claims.
 \item The largest source of systematic errors in the quantities derived in this paper is the lack of uniformity between the sources used for Johnson and
       Str\"omgren photometry. Addressing that issue is one of the objectives of GALANTE, a seven-band photometric survey of the northern Galactic plane we 
       are undertaking using the 80~cm telescope of the Javalambre Astrophysical Observatory in Teruel, Spain. The survey includes a significant fraction of the
       stars in this paper and is being obtained with narrow and 
       intermediate filters specially designed to measure the properties of OB stars. GALANTE is using
       exposure times from 0.1~s to 100~s to obtain S/N $>$ 100 from the brightest stars in our sample to AB magnitudes of 16-17. The large field of view of the
       camera ($1.4\degr\times 1.4\degr$) combined with a pixel size of 0\farcs55 allows for alternative internal and external calibration techniques to
       reduce systematic errors such as zero-point offsets. The survey began in 2016 and, when complete, will be used to revisit the results in this paper.
 \item We will extend this work to larger samples once additional GOSSS spectra are published.
 \item The 2016 (first) Gaia data release included the TGAS parallaxes \citep{Michetal15} but those were of little use to estimate distances to O stars \citep{Maizetal17d}. The second
       Gaia data release is planned for 2018 and should include accurate parallaxes for the majority of the stars in this paper, significantly improving the poor-quality spectroscopic
       parallaxes in Table~\ref{maintable} and facilitating
       exciting new science. 
\end{itemize}

\begin{acknowledgements}
We would like to thank the anonymous referee for a detailed and careful reading of the paper, which led to its significant improvement. We also thank the collaborators who have contributed 
throughout the last decade and a half to the building of the GOSC database that has made this paper possible. In chronological order, those are H. \'A. Galu\'e, L. H. Wei, R. Y. Shida, A. Sota, 
A. T. Gallego Calvente, M. Penad\'es Ordaz, \'A. Alonso Morag\'on, L. Ortiz de Z\'arate Alcarazo, A. Mart\'{\i}n Guti\'errez, and M. Oliva Rubio. All of them heard the lead author say that this 
paper would be published one day and that day has finally arrived, after all the literature was revised, all the archival data were collected, all the spectra observed and reduced, all the calibrations 
checked, all the biases eliminated and the random uncertainties correctly estimated, and all the software written, implemented, tested, and optimized. Sometimes science takes its time to be done 
properly and rushing to publication is not always the best idea. It may collect more citations but ``it ain't right'' 
if it adds more noise than signal to the pool of scientific knowledge.
J.M.A. acknowledges support from the Spanish Government Ministerio de Econom{\'\i}a y Competitividad (MINECO) through grants 
AYA2013-40\,611-P and AYA2016-75\,931-C2-2-P (plus other long expired grants).
R.H.B. acknowledges support from the ESAC Faculty Council Visitor Program, which allowed the two authors to get together for one last time and wrap up the paper.
This research has made use of the SIMBAD database and the VizieR catalog access tool, both operated at CDS, Strasbourg, France.
\end{acknowledgements}

\bibliographystyle{aa}
\bibliography{general}

\begin{appendix}

\section{On the use of filter-integrated or monochromatic quantities for studying extinction}

$\,\!$\indent In previous works (\citealt{Maiz13b}; MA14)  we have briefly discussed the problems associated with the use of 
filter-integrated quantities for studying extinction, an issue that has been known of for a long time \citep{Blan56,Blan57} but frequently ignored. In these
Appendices we provide a more thorough treatment.
Observed photometric magnitudes such as Johnson \UJ \BJ \VJ, Tycho-2 \BT \VT, Str\"omgren \uS \vS \bS \yS, Gaia \GG\, or 2MASS \JT \HT \KT\ are 
filter-integrated quantities, for example, for Johnson \VJ :

\begin{equation}
\VJ = -2.5\log_{10}\left(\frac{\int P_{V_J}(\lambda)f_{\lambda}(\lambda)\lambda\,d\lambda}
                              {\int P_{V_J}(\lambda)f_{\lambda{\rm,Vega}}(\lambda)\lambda\,d\lambda}\right)
                              + {\rm ZP}_{V_J},
\label{VJ}
\end{equation}

\noindent where $P_{V_J}(\lambda)$ is the total-system dimensionless sensitivity function, $f_\lambda(\lambda)$ is the spectral energy
distribution (SED) of the object, $f_{\lambda{\rm,Vega}}(\lambda)$ is the Vega SED, and ZP$_{V_J}$ is the photometric zero-point
\citep{Maiz05b,Maiz06a,Maiz07a}. If the equivalent unextinguished SED, $f_{\lambda,0}(\lambda)$ has a Johnson \VJ\ magnitude of
$V_{J,0}$, then the extinction in that filter, \AVJ, can be expressed as

\begin{equation}
\AVJ \equiv \VJ - V_{J,0} = -2.5\log_{10}\left(\frac{\int P_{V_J}(\lambda)f_{\lambda}(\lambda)\lambda\,d\lambda}
                                                    {\int P_{V_J}(\lambda)f_{\lambda,0}(\lambda)\lambda\,d\lambda}\right).
\label{AVJ}
\end{equation}

An equivalent expression for other magnitudes can be easily written, for example, for Johnson \BJ :

\begin{equation}
\ABJ \equiv \BJ - B_{J,0} = -2.5\log_{10}\left(\frac{\int P_{B_J}(\lambda)f_{\lambda}(\lambda)\lambda\,d\lambda}
                                                    {\int P_{B_J}(\lambda)f_{\lambda,0}(\lambda)\lambda\,d\lambda}\right).
\label{ABJ}
\end{equation}

From the previous equations we arrive at the frequently used definition of (filter-integrated) color excess (or reddening):

\begin{equation}
E(B-V) \equiv \ABJ - \AVJ =  (\BJ - B_{J,0}) - (\VJ - V_{J,0}),
\label{EBV}
\end{equation}

\noindent where, for notation convenience\footnote{We also adopt the usual convention of calling $E(B-V)$ ``the'' color excess.
In reality, one can define a color excess for any combination of two filters e.g., $E(U-B)$.}, we have dropped the $J$ 
subscript in $E(B-V)$. The final filter-integrated quantity of interest is $R_V$, defined as the ratio between total extinction 
in the Johnson's $V$ filter and the color excess:

\begin{equation}
R_V \equiv \frac{\AVJ}{E(B-V)},
\label{RV}
\end{equation}

\noindent where we have also dropped the $J$ subscript in $R_V$ for notation convenience. 

Extinction by dust alters $f_{\lambda,0}(\lambda)$ to yield $f_\lambda(\lambda)$ and is usually expressed in magnitude form as

\begin{equation}
\Al = -2.5\log_{10}\left(\frac{f_{\lambda}(\lambda)}
                              {f_{\lambda,0}(\lambda)}\right).
\label{Al}
\end{equation}

Where \Al\ is the total monochromatic extinction and is a function of the dust properties and of the amount of extinction. It is usually
normalized by the amount of extinction (see below) and in that case it is expressed as \al. That quantity is referred to as the extinction 
law and is a function only of the dust properties\footnote{Here we consider only the ``pure'' extinction case, where radiation originates 
in a single point source and the dust cloud is located far away from both the source and the observer. If one relaxes those assumptions the 
relationship between emitted and observed fluxes depends on radiation transport effects (e.g., scattering back into the beam) and one should talk 
about dust attenuation, not extinction.}.

We can use Eq.~\ref{Al} to express $f_{\lambda}(\lambda)$ as a function of $f_{\lambda,0}(\lambda)$ and insert the result in 
Eqs.~\ref{VJ}-\ref{ABJ}. From there, we can see that Eqs.~\ref{EBV}-\ref{RV} depend not only on \al\ but also on integrals that 
include $f_{\lambda,0}(\lambda)$ and the amount of extinction. In other words, {\bf $E(B-V)$ and $R_V$ are quantities that depend in a complex
way on the amount and type of dust (or extinction) and also on the type of star we are observing.} This is contrary to the common use of 
$E(B-V)$ to linearly characterize the amount of extinction (independent of dust type and SED) and of $R_V$ to characterize the type of dust
(independent of amount of exinction and SED).

In \citet{Maiz04c} we began by
using the monochromatic (or single-wavelength) equivalents to Eqs.~\ref{EBV}-\ref{RV}:

\begin{eqnarray}
\ebv & \equiv & A_{4405} - A_{5495} \nonumber \\
     & =      & [f_{\lambda}(4405) - f_{\lambda,0}(4405)] - \label{ebv} \\
     &        & [f_{\lambda}(5495) - f_{\lambda,0}(5495)], \nonumber
\end{eqnarray}

\begin{equation}
\rv \equiv \frac{A(5495)}{\ebv},
\label{rv}
\end{equation}

\noindent where wavelengths are expressed in \AA. By avoiding the integrals in Eqs.~\ref{AVJ}-\ref{ABJ}, Eq.~\ref{ebv} is a direct and
linear measurement of the amount of extinction (or, more properly, of reddening but for a given extinction law one is a multiple of the other)
and Eq.~\ref{rv} depends only on the type of extinction (or the extinction law). Examples of the differences between $E(B-V)$ and \ebv\ and
between $R_V$ and \rv\ are shown in Fig.~3 of \citet{Maiz13b} for CCM extinction laws (the effect of switching to other families of extinction
laws such as MA14 is small). One can see there that $E(B-V)$ (or, indeed, any other filter-integrated color excess) has a 
non-linear dependence on the amount of extinction. 

The values of 4405~\AA\ and 5495~\AA\ were chosen by \citet{Maiz04c} as representative of the central wavelengths of the \BJ\ and \VJ\ 
filters, respectively, and also to approximately satisfy the limits:

\begin{equation}  
\lim_{\Al\rightarrow 0} E(B-V) \approx \ebv, 
\label{EBVlimit}
\end{equation}  

\begin{equation}  
\lim_{\Al\rightarrow 0} R_V    \approx \rv
\label{RVlimit}
\end{equation}  

\noindent for OB-star SEDs. The choice is reflected in Fig.~3 of \citet{Maiz13b}: in the left panel $R_V \approx \rv$ for 
\teff = 30\,000 K and \ebv = 0 and in the right panel $E(B-V) \approx \ebv$ for \teff = 30\,000 K in the range \ebv = 0-1. We note, however, that
outside those values there are significant differences between the monochromatic and filter-integrated quantities.

For the reasons above, the family of extinction laws of MA14 are parameterized in terms of \ebv\ (amount of extinction) and
\rv\ (type of extinction) instead of $E(B-V)$ and $R_V$. But what about other families? The authors of the CCM and F99 papers either were 
unaware of the issue or did not consider it to be important. In any case, we have used their family of extinction laws by substituting the
filter-integrated quantities by their monochromatic equivalents, as not doing it is [a] unpractical in terms of calculations (the process would 
require several iterations until convergence) and [b] unlikely to have been the authors intentions.

We propose that this issue is so often ignored in the literature for two likely reasons.
The first is that as long as one is working with low-extinction OB stars,
Eqs.~\ref{EBVlimit}-\ref{RVlimit} tell us that substituting the monochromatic quantities by their filter-integrated equivalents can be a good
approximation. The second reason is that filter-integrated quantities can be easily computed from the observed photometry and spectral types while
monochromatic quantities have to be calculated using additional information and numerical techniques. Indeed, those two reasons may have been
valid 40 or 50 years ago, when most targets of extinction studies were low-extinction OB stars and computing facilities were limited. However,
today they are not valid excuses as massive photometric and spectroscopic studies provid us with data for stars of all spectral types 
and for extinctions that probe into much larger Galactic distances. Confusing the two types of quantities can easily lead (as it is too often the
case) to biases in photometric measurements of extinction.

\section{Limitations of the $Q$ approximation}

$\,\!$\indent The $Q$ approximation is a method used to determine extinctions of early-type stars that dates back to
\citet{JohnMorg53}. $Q$ is a linear combination of $(U-B)$ and $(B-V)$ of the form:

\begin{equation}
Q = (U-B) - \alpha(B-V),
\label{Q1}
\end{equation}

\noindent where $\alpha$ is a constant for which \citet{JohnMorg53} give a value of 0.72$\pm$0.03. They claimed that $Q$ is nearly independent
of extinction and can be used to determine the spectral type of the target, giving values that start at $-0.93$ for most O stars (except the 
later types), $-0.70$ for B2, $-0.44$ for B5, and $0.00$ for A0\footnote{These are the values calculated by \citet{JohnMorg53} and they 
can differ from those calculated here for the examples below.}. As $Q$ can be used as a proxy for the spectral type, the color excess can 
be also calculated in a second step. \citet{JohnMorg53} give:

\begin{equation}
E(B-V) = (B-V) - 0.337Q + 0.009.
\label{Q2}
\end{equation}

In its original form or in adapted versions, the $Q$ approximation is used even today due to its simplicity. Nevertheless, it has some known 
limitations: 

\begin{itemize}
 \item Unless some additional information is present, it is restricted to OB stars. For A stars and later types, solutions become multiple as,
       for example, a low-extinction A star can be confused with a higher-extinction late-B star \citep{Maiz04c}. 
 \item The value of $\alpha$ depends on the extinction law for example, \rv, so it is not a real constant. \citet{JohnMorg53} did their analysis 
       assuming an average extinction law but for non-canonical values of \rv\ one needs to compute the corresponding $\alpha$ (see below for
       examples). With such an additional parameter, the application of the $Q$ approximation is not straightforward.
 \item For late-B stars (and later types) two stars of the same spectral types but different luminosity classes can have significantly 
       different intrinsic Johnson colors, thus complicating the method.
 \item Emission-line stars need to be excluded beforehand in order to avoid biased results.
 \item The $Q$ approximation can be adapted to similar filter sets (e.g., HST/WFPC2 F336W+F439W+F555W) but the equivalent values of $\alpha$ 
       have to be computed.
\end{itemize}

To those limitations one has to add the non-linearity of colors described in the previous Appendix: $(U-B)$ and $(B-V)$ do not increase with
the amount of extinction at a constant rate, as the rate depends on the amount of extinction and on the spectral type. Therefore, in an
$(U-B)$ vs. $(B-V)$ color plane, the trajectories created by increasing extinction are not straight lines (as Eq.~\ref{Q1} indicates) and
are not parallel for the same amount of extinction if the spectral type is different. The reader is referred to Fig.~1 of \citet{Maiz04c} 
for a graphical representation of the effect. 

Even though knowledge of those limitations should be widely known, unfortunately one still sees papers where the $Q$ approximation is
incorrectly applied and, as a result, the published extinction results are biased. To show the relevance of the effects described above, we
have used the SED grid of \citet{Maiz13a} and the MA14 family of extinction laws to compute the real value of $\alpha$ for different amounts 
and types of extinction as well as input SEDs (Table~\ref{alpha}). In that way we can quantify some of the effects described above:

\input{table_ap1}

\begin{itemize}
 \item The first column in Table~\ref{alpha} lists $\alpha$ for low canonical extinction applied to MS stars, in other words, the case 
       that was likely considered by \citet{JohnMorg53}. For \teff $\ge$ 10\,000~K, the classical value of 0.72 is not a poor approximation, but 
       there is considerable scatter (from 0.709 to 0.780) and the average is above 0.72 (see the cases below for possible 
       explanations of the latter effect). For lower values of \teff, $\alpha$ can be significantly higher already in this regime. 
 \item The next two columns show increasingly larger values of the canonical extinction applied to MS stars. Already for \ebv\ = 1.0 we see how
       non-linear effects increase the value of $\alpha$ to the point of making it larger than 0.72 for all MS OB stars (and much larger for 
       cool stars). For \ebv\ = 3.0 $\alpha$ is considerably higher and the simple $Q$ approximation fails completely. 
 \item A comparison between the second and fourth columns in Table~\ref{alpha} shows the effect of changing \rv\ from 3.1 to 5.0. The increase
       in \rv\ reduces the value of $\alpha$. It is possible that the original \citet{JohnMorg53} analysis included some objects with \rv\
       above the canonical value that decreased their average $\alpha$.
 \item Finally, a comparison between the first and fifth columns shows the effect of luminosity class (or gravity). Differences are negligible
       for O and early-B stars but become appreciable for late-B stars and very large for cool stars. Late-B supergiants have lower values of
       $\alpha$ than their MS counterparts, which could also help explain the average 0.72 value of \citet{JohnMorg53}. 
\end{itemize}

We conclude that {\bf the $Q$ approximation can only be applied (and with care) to OB stars with low canonical extinction.}
In other cases it yields biased results both for \ebv\ and \teff. On the other hand, the use of a code that combines optical and NIR
information handling extinction properly (such as CHORIZOS) can accurately determine \teff\ from photometric data \citep{MaizSota08}, as shown 
by MA14. {\bf It may have been reasonable to use such an approximation decades ago but with the current computing capabilities 
and extensive availability of well calibrated NIR data it has ceased to be so in most cases.} As this paper shows, it is possible to avoid 
biases and, at the same time, compute \ebv\ and \rv\ simultaneously and detect cases with extreme anomalous SEDs caused by, for example, IR excesses.

\section{Anticorrelation between \ebv\ and \rv}

$\,\!$\indent When using optical+NIR photometry to simultaneously fit \ebv\ and \rv\ to a star with a known unextinguished SED, one obtains
uncertainties for both quantities, that is, $\sigma_{E(4405-5495)}$ and $\sigma_{R_{5495}}$. However, the final objective of the process is usually to 
obtain the extinction in a given band (e.g., \AVJ) or the unextinguished magnitude (e.g., $V_{J,0}$), so the fitted quantities are intermediate
values. One may be tempted to obtain \AVJ\ by multiplying \ebv\ and \rv\ and to obtain its relative uncertainty ($\sigma_{A_V}$, dropping the 
$J$ to avoid a triple subscript) by summing in quadrature the relative uncertainties for \ebv\ and \rv. This is wrong for two reasons. 
The first is that:

\begin{equation}
A_{5495} = \ebv\, \rv \ne \AVJ,
\label{A5495}
\end{equation}

\noindent as $A_{5495}$ is a monochromatic quantity and \AVJ\ a filter-integrated one. The second reason is that the two fitted quantities,
\ebv\ and \rv, can be correlated so the propagation of uncertainties yields:

\begin{eqnarray}
 \left(\frac{\sigma_{A_{5495}    }}{A_{5495}}\right)^2  & = &
 \left(\frac{\sigma_{E(4405-5495)}}{\ebv    }\right)^2 + 
 \left(\frac{\sigma_{R_{5495}    }}{\rv     }\right)^2 + \label{sA5495} \\
 & &
2\left(\frac{\sigma_{E(4405-5495)}}{\ebv    }\right)
 \left(\frac{\sigma_{R_{5495}    }}{\rv     }\right) \rho_{E(4405-5495),R_{5495}}, \nonumber
\end{eqnarray}

\noindent where $\rho_{E(4405-5495),R_{5495}}$ is the Pearson correlation coefficient between the two quantities. In principle, 
$\rho_{E(4405-5495),R_{5495}}$ could be of either sign but, as we show in the next Appendix, it is always negative and large under reasonable 
assumptions, in other words, \ebv\ and \rv\ are strongly anticorrelated in most situations.

\begin{figure}
\centerline{\includegraphics[width=\linewidth]{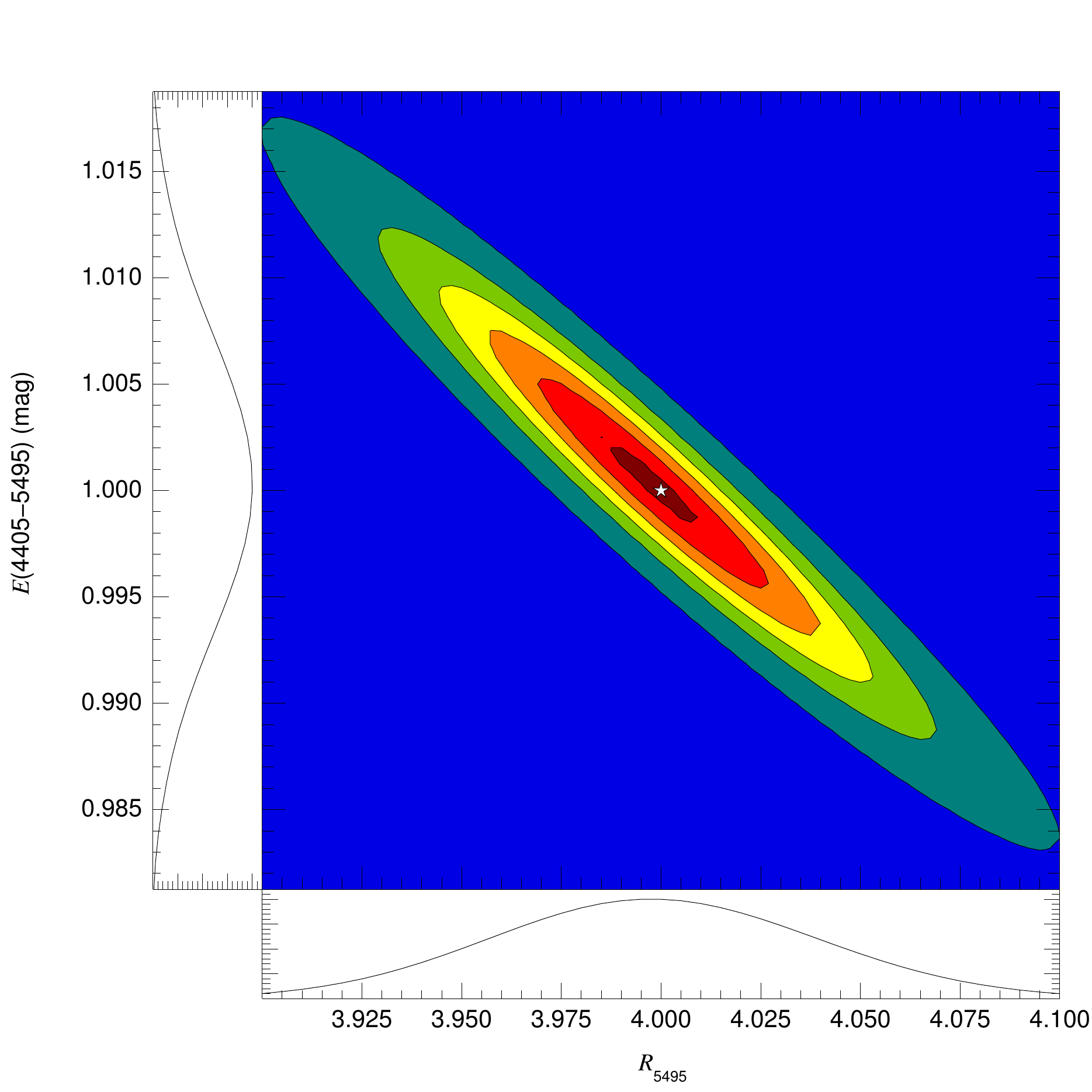}}
\caption{\rv-\ebv\ plane likelihood for a simulated CHORIZOS run. The input \UJ \BJ \VJ \JT \HT \KT\ photometry corresponds to a SED with 
         input \teff\ = 30\,000 K, LC = 5, MW-metallicity, \ebv\ = 1.0, \rv\ = 4.0 at a distance of 10 pc with uncertainties of 0.01 mag for each 
         filter. The output has \ebv\ = 1.0003$\pm$0.0067, \rv\ = 3.999$\pm$0.039, $A_{5495}$ = 4.0002$\pm$0.0151, and \AVJ\ = 4.0313$\pm$0.0151 
         (see Eq.~\ref{A5495}). We note how the relative uncertainties of $A_{5495}$ and \AVJ\ (0.38\%) are smaller than either that of \ebv\ 
         (0.67\%) or \rv\ (0.98\%) due to the strong anticorrelation ($\rho_{E(4405-5495),R_{5495}} = -0.962$) between the latter two quantities.}
\label{anticorr}
\end{figure}

Figure~\ref{anticorr} shows the CHORIZOS likelihood output for an example built from input synthetic photometry. As expected, $A_{5495} \ne \AVJ$,
though the effect is small in this particular case (it becomes larger for higher extinctions or for later types). On the other hand, the effect in 
$\sigma_{A_V}$ is very large due to the high value of $\rho_{E(4405-5495),R_{5495}}$. This example is representative of the sample in this paper,
indicating that {\bf the anticorrelation between \ebv\ and \rv\ is an important effect that should be taken into account when computing 
uncertainties for optical extinctions.}

We now turn our attention to an explanation behind this anticorrelation. One way to look at it is that if, for a given unextinguished SED, one knows \VJ\ and \KT, then 
\AVJ\ is relatively well determined because the \KT\ magnitude ``anchors'' the total extinction reasonably well, as that filter is little affected
(in relative terms) by the presence of dust along the light path. How to fit the additional photometric points is a different story, as it becomes 
possible to vary \ebv\ and \rv\ in synchrony maintaining a fixed total extinction. This is the kind of effect that a numerical code that processes 
multifilter photometry simultaneously can find that simpler methods such as the $Q$ approximation cannot.

An additional related issue is how to obtain $V_{J,0}$ and its uncertainty. One could simply calculate $\VJ-\AVJ$ and sum the uncertainties in
quadrature. Alternatively, one can use the whole likelihood grid calculated by CHORIZOS (2-D if \ebv\ and \rv\ are the only quantities being fitted or 
of higher dimensions if other quantities such as \logd\ or \teff\ are included in the fit) and derive $V_{J,0}$ directly by summing over the whole grid. 
The second method (which is the one used in Table~\ref{maintable}) has two advantages: [a] it uses information from all the photometric points and 
[b] it naturally takes into account the possible anticorrelation between \VJ\ and \AVJ\ (as a dim \VJ\ can be caused by a large \AVJ). Both of those 
advantages tend to lower the uncertainties. On the other hand, the second method has the disadvantage of being model-dependent and, hence, more prone
to be affected by systematic uncertainties (caused by e.g., stars with strong emission lines). It is easy to set up an example where the second method 
improves the value of $V_{J,0}$ over the first one thanks to the use of information from all photometric points: for a low-extinction SED where one 
has both Johnson and Tycho-2 photometry and $\sigma_{V_J} \gg \sigma_{V_T}$, then the information contained on \VT\ reduces $\sigma_{V_{J,0}}$
considerably\footnote{But we urge the reader not make the mistake of doing \VJ\ = \VT: the transformation from one magnitude to another always involves one color. 
The color may be close to zero for some \teff\ and extinctions and significant for others.}

\section{Uncertainties in extinction as a function of photometric uncertainties}

$\,\!$ \indent To test that CHORIZOS yields the correct values for the extinction parameters, we set up the following experiment:

\begin{itemize}
 \item For a star with a known unextinguished SED, we computed \UJ \BJ \VJ \JT \HT \KT\ for a range of \ebv\ values between 0.05 and 1.00 and
       \rv\ values between 3.0 and 7.0 using the MA14 familiy of extinction laws.
 \item We assigned the same photometric uncertainty to all filters 
       (e.g., $\sigma_{U_J}=\sigma_{B_J}=\sigma_{V_J}=\sigma_{J}=\sigma_{H}=\sigma_{K_{\rm s}}=0.01$ mag) and we randomly changed each magnitude 
       according to a Gaussian distribution with that uncertainty. The process was repeated using different Montecarlo realizations and different 
       photometric uncertainties.
 \item We processed each of the simulated objects through CHORIZOS fitting \ebv, \rv, and \logd\ simultaneously while leaving the rest of the parameters
       fixed (to their known values).
\end{itemize}

The fitted \ebv\ and \rv\ and their uncertainties ($\sigma_{E(4405-5495)}$ and $\sigma_{R_{5495}}$) were compared with the input values and the 
results were excellent:

\begin{itemize}
 \item No significant bias is present. When a large number of Montecarlo realizations with the same input parameters were processed through 
       CHORIZOS, the difference between the mean of the fitted values and the input was always much less than the calculated
       uncertainties.
 \item The calculated uncertainties are consistent with the dispersion observed in the Montecarlo realizations. In other words, combining this
       point and the previous one, {\bf CHORIZOS correctly calculates random uncertainties without introducing systematic errors.}
 \item The calculated uncertainties follow functional forms that allow approximate analytical results to be derived:
\end{itemize}

\begin{eqnarray}
\sigma_{E(4405-5495)}                        & \approx & 0.70 \, \sigma_{V_J}                         \label{sigmaebv} \\
\sigma_{R_{5495}}                            & \approx & 0.68 \, \sigma_{V_J} \, (\rv+2) \, / \, \ebv \label{sigmarv}  \\
\sigma_{A_{5495}} \, \approx \, \sigma_{A_V} & \approx & 1.58 \, \sigma_{V_J}                         \label{sigmaav} 
\end{eqnarray}

From Eqs.~\ref{anticorr}~and~\ref{sigmaebv}-\ref{sigmaav} we obtain:

\begin{equation}
\rho_{E(4405-5495),R_{5495}} \approx - \frac{\mbox{$R_{5495}^2$} + 1.942\,\rv\, - 0.681}{\mbox{$R_{5495}^2$} + 2\,\rv},
\label{rho}
\end{equation}

\noindent which, for \rv\ in the range from 3 to 7, varies between $-0.943$ and $-0.983$ (i.e., strongly anticorrelated, as anticipated in the
previous Appendix). 

One important consequence of Eq.~\ref{sigmarv} is that there is a limit to how well can \rv\ be measured that depends strongly on \ebv. For
example, if $\sigma_{V_J}$ = 0.03, an extinction of \ebv\ = 0.10 and \rv\ = 4.0 allows one to measure the latter with an uncertainty of 1.2, which 
is too large to yield much information. We can qualitatively describe what is going on: it is easy to measure the properties of an
effect when the effect is large but difficult when the effect is small. It is also important to keep in mind that, unfortunately, when working with 
real-world photometry from multiple sources there are usually hidden calibration issues (e.g., \citealt{Maiz05b,Maiz06a,Maiz07a}) plus possible
disparities between the real and model SEDs (emission lines, IR excesses\ldots) that introduce systematic errors. Therefore, {\bf one should avoid 
low-extinction objects when calculating optical-NIR extinction properties.} 

We would like to end with a cautionary note on the value of Eqs.~\ref{sigmaebv}-\ref{rho}. They have been calculated for the specific experiment
described here. We would expect that if the experiment conditions are somewhat altered (different filters, known distance, unknown \teff, non-uniform 
photometric uncertainties\ldots) the qualitative behavior should be similar but the quantitative one (i.e., coefficients) should change (e.g., see 
Fig.~\ref{ebv_srv} where increasing the number of filters decreases the value of $\sigma_{R_{5495}}$). 

\section{On the FM09 family of extinction laws}

$\,\!$ \indent It was not our original intention to comment on the applicability of the \citet{FitzMass09} or FM09 family of extinction laws
to our data but the referee requested that we explain why it was not included along with CCM and F99 in our list of comparison families and so we 
discuss it in this Appendix.

In FM09, ACS/HST spectrophotometry was used in the 6000-9500~\AA\ range combined with \VJ\ and 2MASS~\JT\HT\KT\ photometry to derive a new family of extinction 
laws of the form (Eq.~5 in their paper):

\begin{equation}
\frac{\Al-\AVJ}{E(B-V)} = \frac{0.349 + 2.087 R_V}{1 + (\lambda/0.507)^\alpha} - R_V,
\label{FM09_01}
\end{equation}

\noindent where $\lambda$ is in microns and we have adapted their notation to the one in this paper. The FM09 family of extinction laws is different from
the CCM. F99, or MA14 ones in that it is a two-parameter family ($R_V$ and $\alpha$) instead of a one-parameter family. The introduction of a second 
parameter should ease the fitting of a larger number of photometric points but, of course, it should be used only when necessary, as Occam's razor 
dictates. FM09 use as an argument for doing so their previous results of \cite{FitzMass07}, their paper V (FM09 is their paper VI). 

We can use Eq.~\ref{RV} to rearrange Eq.~\ref{FM09_01} into a more convenient form:

\begin{equation}
\frac{\Al}{\AVJ} = \frac{0.349 + 2.087 R_V}{R_V} \frac{1}{1 + (\lambda/0.507)^\alpha}, 
\label{FM09_02}
\end{equation}

\noindent where we have divided the right side of the equation into two different terms to study their effect on the left side. The first term depends
only on $R_V$ (the first parameter), so it is a constant for a given choice of the extinction law. The second term contains the dependence on $\lambda$
as well as that on $\alpha$ (the second parameter). 

The first issue with Eq.~\ref{FM09_02} we have dealt with in Appendix A: \AVJ\ and $R_V$ are filter-integrated quantities that cannot be
properly used to characterize an extinction law, as they depend in a complex way on the amount and type of extinction and also on the SED we are
observing. Therefore, we need to find the equivalent to Eq.~\ref{FM09_02} using monochromatic quantities. We can simply substitute \AVJ\ by
$A_{5495}$ but in the case of $R_V$ we make the substitution $f = (0.349 + 2.087 R_V)/R_V$ for reasons that will become clear later. Therefore, we end
up with:

\begin{equation}
\frac{\Al}{A_{5495}} = f \frac{1}{1 + (\lambda/0.507)^\alpha}. 
\label{FM09_03}
\end{equation}

Once we have reached this point, there is one point that is unclear in FM09: what is the range of applicability of Eq.~\ref{FM09_03}? Does it extend to shorter
wavelengths than 6000~\AA\ (e.g., to the whole optical-NIR range) or is it constrained to $\lambda > 6000$~\AA? We consider both of those options below.

{\em Option 1: Equation~\ref{FM09_03} extends to the whole optical-NIR range.} If this is the case, we can plug in $\lambda$~=~5495~\AA~=~0.5495~$\mu$m
to Eq.~\ref{FM09_03}. As the left hand side has to be equal to one, then:

\begin{equation}
f = 1 + (0.5495/0.507)^\alpha = 1 + 1.0838^\alpha,
\label{fvalue}
\end{equation}

\noindent which leads to:

\begin{equation}
\frac{\Al}{A_{5495}} = \frac{1 + 1.0838^\alpha}{1 + (\lambda/0.507)^\alpha}. 
\label{FM09_04}
\end{equation}

We note that Eq.~\ref{FM09_04} is independent of \rv\ (or of $R_V$) that is, it is actually a single-parameter family of extinction laws, as the only parameter
left in the equation is $\alpha$. Furthermore, the definition of $f$ and Eq.~\ref{fvalue} lead to:

\begin{equation}
R_V = \frac{0.349}{f-2.087} = \frac{0.349}{1.0838^\alpha-1.087},
\label{rvvalue1}
\end{equation}

\noindent that is, $R_V$ is a function of $\alpha$, not an independent parameter. We can also plug in $\lambda$~=~4405~\AA~=~0.4405~$\mu$m in 
Eq.~\ref{FM09_03} and use Eqs.~\ref{rv}~and~\ref{fvalue} to arrive at:

\begin{equation}
\rv = \frac{1+0.8688^\alpha}{1.0838^\alpha-0.8688^\alpha}.
\label{rvvalue2}
\end{equation}

Equations~\ref{rvvalue1}~and~\ref{rvvalue2} are incompatible. For example, for $\alpha = 1.6$ the first one yields 6.92 and the second one 5.31 while
for $\alpha = 2.9$ the first one yields 1.99 and the second one 2.79. We note that changing the central wavelengths of the \BJ\VJ\ filters to values other
than 4405~\AA\ and 5495~\AA\ changes the coefficients in Eqs.~\ref{rvvalue1}~and~\ref{rvvalue2} but not the functional forms, which are incompatible. 

This reasoning leads us to the conclusion that option 1 is not what the authors of FM09 must have intended and therefore we discard it. We note, however, that
other authors may be unaware of this \footnote{We do not think
we are the first to notice this issue with FM09. \citet{Schletal16} correctly indicate 
at one point that ``FM09 has a second parameter, $R_V$, that does not change the shape of the extinction curve redwards of $V$'' and later on mention that 
``we have excluded the $g_{\rm P1}$ band, which is outside the region where the FM09 prescription was developed'' (note that \citet{Schletal16} does 
include $r_{\rm P1}$ in their FM09 comparison even though that filter extends redward of 6000~\AA). If one adds those two statements together, those authors 
are saying that FM09 is not applicable to the \BJ\ and \VJ\ bands while $R_V$ has no effect in its range of applicability. Therefore, their conclusion must 
be the same as ours: the presence of $R_V$ as an independent parameter in FM09 is spurious.}.

{\em Option 2: Equation~\ref{FM09_03} is applicable only to $\lambda > 6000$~\AA.} In this case, the authors would have implied that a different extinction law is 
required blueward of that wavelength. However, that extinction law is not explicitly mentioned anywhere in their paper. Since the majority of the photometric 
data in our paper is in the 3000-6000~\AA\ range, we are unable to apply the FM09 family of extinction laws for our data. Nevertheless, let us explore the consequences 
of the FM09 family of extinction laws redwards of 6000~\AA.

We start by noting that if a different extinction law has to be used, then Eq.~\ref{fvalue} is not strictly true, as \Al/$A_{5495}$ has a different (not specified)
form around that wavelength. However, extinction curves are continuous and differentiable and since 5495~\AA\ is close in wavelength to 6000~\AA, then Eq.~\ref{fvalue} 
has to be approximately valid. We show that is indeed the case in Table~\ref{FM09_table}, where we extract the results of Table~4 of FM09 and compute the first term of 
Eq.~\ref{FM09_02} and the inverse of the second term of that same equation evaluated at 5495~\AA. As it can be seen, the values are nearly identical (in most cases 
the differences can be adscribed to round-off issues in the FM09 table). The conclusion from this experiment is that, effectively, FM09 is a single-parameter family
of extinction laws, not a two-parameter family as claimed, as $R_V$ is completely determined by $\alpha$. Therefore, Eq.~\ref{FM09_04} is a valid parametrization for
FM09 redwards of 6000~\AA\ (we suspect this formula or a similar one is what \citealt{Schletal16} used, see footnote above). An interesting corollary is that for 
the blueward extension of FM09 to be consistent in the sense of $R_V \approx \rv$, its functional form would have to be well constrained by the redwards portion (and 
different from Eq.~\ref{FM09_04}), so even then it would be a quasi-single-parameter family.

\begin{table}
\caption{Transformation of the results from Table~4 of FM09 using the functions described in the text.}
\label{FM09_table}
\begin{tabular}{lcc}
Star & $(0.349+2.087 R_V)/R_V$ & $1 + 1.0838^\alpha$ \\
\hline
BD +56 517     & 2.209 & 2.262 \\
BD +45 973     & 2.219 & 2.222 \\
BD +44 1080    & 2.239 & 2.220 \\
NGC 1977-885   & 2.150 & 2.149 \\
HD 46\,106     & 2.213 & 2.218 \\
HDE 292\,167   & 2.202 & 2.213 \\
HD 68\,633     & 2.186 & 2.227 \\
HD 70\,614     & 2.203 & 2.221 \\
Trumpler 14-6  & 2.160 & 2.169 \\
Trumpler 14-27 & 2.165 & 2.157 \\
Herschel 36    & 2.145 & 2.138 \\
HDE 229\,196   & 2.200 & 2.207 \\
HD 204\,827    & 2.236 & 2.248 \\
BD +61 2365    & 2.199 & 2.201 \\
\hline
\end{tabular}
\end{table}

This does not necessarily mean that the results of FM09 are invalid. What those authors could have derived is a single-parameter family of extinction laws valid 
for $\lambda > 6000$~\AA\ that behaves as a power law in the long-wavelength regime and becomes modified in the optical, an idea that they adopted from a suggestion
by Karl Gordon (see their footnote 6). However, before accepting the FM09 results we must end with three additional words of caution.

Firstly, the authors state that ``We thus quadratically combine the standard 2MASS uncertainties with values of 0.007, 0.040, and 0.017 mag for \JT, 
\HT, and \KT, respectively, to arrive at the total uncertainty for each magnitude''. In this paper we did not require such additional uncertainties (except for objects with
bad quality flags), which leads us to think that 2MASS photometry is correct and that there must be a source of uncertainties unaccounted for in FM09.

Secondly, the FM09 sample consists only of 14 objects. We think that number is too small to properly constraint a family of extinction laws, as it does not
provide an adequate sampling of the parameter space. Furthermore, some of the stars in their sample have quite a low extinction, making the IR measurements suspect.
For example, for HD~46\,106 we derive $\ebv = 0.392\pm0.007$ and the spectral type is O9.7~III(n), not B1~V as stated in FM09. Also, only two stars in their Table~1 are
listed as having $E(B-V) > 1.0$.

Finally, the object that FM09 chose as their highest-\rv\ representative is Herschel~36 (a.k.a. HD~164\,740), a star to which they pay special attention. However, the 
authors are apparently unaware of the work of \citet{Gotoetal06}, who showed that an obscured companion has a significant contribution in the NIR, or of \citet{Ariaetal06},
whose Fig.~2 clearly showed the NIR excess in comparison with the primary SED, in both cases three years prior to the publication of their paper (see also the discussion 
regarding Herschel~36
in the main body of this paper). The presence of such anomalous objects
in a sample is the reason why any target list used for extinction law calculations has to be carefully culled, as we have done in this paper.

\end{appendix}

\end{document}

%% file: maintable.tex
\longtabL{1}{
\begin{landscape}
\begin{longtable}{lcclr@{$\pm$}lr@{$\pm$}lr@{$\pm$}lr@{$\pm$}lcrrr}
\caption{\label{maintable} 
          Results from the CHORIZOS fits. The name includes all GOSSS components (the NC column gives 
          the number of components) while the GOSSS ID refers to the brightest component when there are two.
          The SS column gives the subsample to which the object belongs. The results for \ebv, \rv, \AVJ, 
          $V_{J,0}$, and \logd\ are for the MA14 family of extinction laws.}\\
\hline\hline
Name & GOSSS ID & NC & SS & \multicolumn{2}{c}{\ebv}  & \multicolumn{2}{c}{\rv} & \multicolumn{2}{c}{\AVJ}  & \multicolumn{2}{c}{$V_{J,0}$} & \logd & \chir & \chir & \chir \\
     &          &    &    & \multicolumn{2}{c}{(mag)} & \multicolumn{2}{c}{}    & \multicolumn{2}{c}{(mag)} & \multicolumn{2}{c}{(mag)}     & (pc)  & CCM   & F99   & MA14  \\
\hline
\endfirsthead
\caption{continued.}\\
\hline\hline
Name & GOSSS ID & NC & SS & \multicolumn{2}{c}{\ebv}  & \multicolumn{2}{c}{\rv} & \multicolumn{2}{c}{\AVJ}  & \multicolumn{2}{c}{$V_{J,0}$} & \logd & \chir & \chir & \chir \\
     &          &    &    & \multicolumn{2}{c}{(mag)} & \multicolumn{2}{c}{}    & \multicolumn{2}{c}{(mag)} & \multicolumn{2}{c}{(mag)}     & (pc)  & CCM   & F99   & MA14  \\
\hline
\endhead
\hline
\endfoot
HD 164\,019             & GOS 001.91$-$02.62\_01 & 1 & J2GT  & 0.495&0.012 & 3.139&0.106 & 1.578&0.023 &  7.706&0.020 & 3.5 &  1.57 &  1.26 &  1.45 \\
63 Oph                  & GOS 004.54$+$00.30\_01 & 1 & J2GTS & 0.300&0.006 & 4.299&0.124 & 1.307&0.016 &  4.880&0.015 & 3.2 &  1.86 &  2.21 &  0.69 \\
HD 168\,941             & GOS 005.82$-$06.31\_01 & 1 & J2GTS & 0.310&0.007 & 3.764&0.117 & 1.182&0.018 &  8.173&0.016 & 3.6 &  0.61 &  0.55 &  0.22 \\
HD 164\,536             & GOS 005.96$-$00.91\_01 & 1 & J2GT  & 0.244&0.010 & 4.218&0.243 & 1.043&0.026 &  6.081&0.024 & 3.1 &  0.99 &  1.36 &  0.57 \\
Herschel 36             & GOS 005.97$-$01.17\_01 & 1 & J2G   & 0.825&0.035 & 5.573&0.293 & 4.616&0.075 &  5.683&0.057 & 3.0 &  0.12 &  0.61 &  0.35 \\
9 Sgr AB                & GOS 006.01$-$01.20\_01 & 1 & J2GTS & 0.315&0.006 & 4.199&0.122 & 1.338&0.021 &  4.620&0.020 & 2.9 &  2.41 &  1.88 &  0.87 \\
HD 164\,816             & GOS 006.06$-$01.20\_01 & 1 & J2GT  & 0.258&0.010 & 4.088&0.208 & 1.068&0.019 &  6.014&0.017 & 3.0 &  1.23 &  1.12 &  0.87 \\
HD 165\,052             & GOS 006.12$-$01.48\_01 & 1 & J2GTS & 0.392&0.006 & 3.828&0.093 & 1.522&0.020 &  5.341&0.019 & 3.0 &  2.99 &  2.02 &  1.11 \\
\end{longtable}
\end{landscape}
}

%% file: clusassoc.tex
\begin{table}
\caption{Number of objects and ranges of \ebv\ and \rv\ for some of the clusters and associations in this paper.}
\label{clusassoc}
\centerline{
\begin{tabular}{lrcc}
Cluster/association    & $N$ & \ebv      & \rv     \\
\hline
M8                     &   4 & 0.26-0.83 & 3.8-5.6 \\
M16                    &   5 & 0.68-1.85 & 3.5-3.9 \\
NGC 6604               &   5 & 0.89-1.14 & 3.3-3.7 \\
Cygnus OB2             &  33 & 1.48-2.88 & 2.6-3.4 \\
Berkeley 59            &   4 & 1.36-2.23 & 2.8-3.3 \\
Heart and Soul Nebulae &  11 & 0.58-0.93 & 3.2-3.7 \\
NGC 1893               &   5 & 0.50-0.79 & 3.0-3.6 \\
NGC 2244               &   6 & 0.39-0.50 & 3.2-3.4 \\
Orion OB1              &   9 & 0.02-0.29 & 4.2-6.4 \\
$\;\;$Orion Nebula     &   2 & 0.20-0.29 & 6.2-6.2 \\
Carina Nebula          &  64 & 0.20-1.19 & 3.0-6.1 \\
NGC 3603               &   4 & 1.19-1.41 & 3.9-4.0 \\
IC 2944                &  10 & 0.27-0.46 & 3.3-3.9 \\
NGC 6231               &  12 & 0.38-0.51 & 3.3-4.1 \\
Havlen-Moffat 1        &   8 & 1.78-2.10 & 2.7-3.2 \\
Pismis 24              &   7 & 1.66-1.91 & 3.2-3.6 \\
\hline
\end{tabular}
}
\end{table}

%% file: table_ap1.tex
\begin{table}
\centerline{
\begin{tabular}{lcccccr}
\hline
\ebv & 0.25  & 1.00  & 3.00  & 1.00  & 0.25  &            \\
LC   & MS    & MS    & MS    & MS    & SG    &            \\
\rv  & 3.1   & 3.1   & 3.1   & 5.0   & 3.1   & \teff\ (K) \\
\hline
     & 0.924 & 0.961 & 1.070 & 0.880 & 1.130 & 3500       \\
     & 0.794 & 0.819 & 0.914 & 0.759 & 0.699 & 7000       \\
     & 0.709 & 0.728 & 0.816 & 0.677 & 0.676 & 10\,000    \\
     & 0.764 & 0.781 & 0.861 & 0.731 & 0.767 & 20\,000    \\
     & 0.780 & 0.797 & 0.874 & 0.748 & 0.780 & 40\,000    \\
\hline
\end{tabular}
}
\caption{Values of $\alpha$ as a function of \teff\ calculated from synthetic photometry for different
         values of \ebv, luminosity class (main sequence, MS, or supergiant, SG), and \rv.}
\label{alpha}
\end{table}